\documentclass[12pt,preprint]{emulateapj}

\usepackage{lscape}
\usepackage{amsmath}

\slugcomment{submitted to ApJ}

\shorttitle{Chandra observations of the galaxy group AWM~5}
\shortauthors{Baldi et al.}

\begin{document}

\title{Chandra observations of the galaxy group AWM~5: 
cool core re-heating and thermal conduction suppression}


\author{A. Baldi, W. Forman, C. Jones, P. Nulsen, L. David, R. Kraft}
\affil{Harvard-Smithsonian Center for Astrophysics}






\and

\author{A. Simionescu}
\affil{Max Planck Institute for Extraterrestrial Physics}






\begin{abstract}
We present an analysis of a 40~ksec Chandra observation of the 
galaxy group AWM~5. 
It has a small ($\sim8$ kpc) dense cool core with a temperature
of $\sim1.2$ keV and the temperature profile 
decreases at larger radii, from $\sim3.5$ keV just outside the core 
to $\sim2$ keV at $\sim300$ kpc from the center.
The abundance distribution shows a ``hole'' in the central $\sim10$~kpc, where
the temperature declines sharply. An abundance
of at least a few times solar is observed $\sim15-20$ kpc from
the center.
The deprojected electron density profile shows a break in slope at $\sim13$~kpc
and can be fit by two $\beta$-models, with $\beta=0.72_{-0.11}^{+0.16}$ and 
$r_c=5.7_{-1.5}^{+1.8}$~kpc, for the inner part, and $\beta=0.34\pm0.01$ and 
$r_c=31.3_{-5.5}^{+5.8}$ kpc, for the outer part.
The mass fraction of hot gas is fairly flat in the center and increases for $r>30$~kpc up 
to a maximum of $\sim6.5\%$ at $r\sim380$~kpc.
The gas cooling time within the central 30 kpc is smaller than a Hubble time,
although the temperature only declines in the central $\sim8$ kpc region. 
This discrepancy suggests that an existing cooling core has been partially
re-heated. In particular, thermal conduction could have been a significant source of re-heating.
In order for heating due to conduction to balance cooling due to 
emission of X-rays, the conductivity must be suppressed by a large factor (at least
$\sim100$).
Past AGN activity (still visible as a radio source in the center of the group) 
is however the most likely source that re-heated the central regions of AWM~5. 
We also studied the properties of the ram pressure stripped tail in the group member NGC~6265.
This galaxy is moving at $M\approx3.4_{-0.6}^{+0.5}$ ($v\sim2300$~km s$^{-1}$) 
through the hot group gas. The physical length of the tail is $\sim42$~kpc and its 
mass is $2.1\pm0.2\times10^9$~M$_\odot$.
\end{abstract}





\section{Introduction}\label{intro}

The galaxy group AWM~5 (Albert et al. 1977) is an X-ray 
luminous nearby group ($z=0.0348$)
with $L_X\sim6\times10^{43}$ erg s$^{-1}$
(Jones \& Forman 1999). The cD galaxy NGC~6269 lies 
at the group center 
and at the peak of the X-ray surface brightness distribution. NGC~6269 contains 
a weak central radio source ($\sim27$ mJy) with north-south radio 
lobes ($\sim8$ and $\sim10$ mJy) extending 
25$^{\prime\prime}$ from the center (Burns et al. 1981, Giacintucci et al. 2007).
NGC~6269 dominates the optical light from AWM~5. Since the group member NGC~6265 is 
only $\sim1.3$ mags fainter than NGC~6269 and is located at $\sim260$~kpc (in the plane
of the sky) from the center, AWM~5 does not strictly follow
the criteria of Jones et al. (2003) for the definition of ``fossil'' groups (e.g. Ponman 
et al. 1994, Mulchaey \& Zabludoff 1999, Jones et al. 2000). 
However projection effects may increase in principle the distance between the two galaxies
pushing it over the half a virial radius limit. Moreover the velocity difference 
(perpendicularly to the plane of the sky) between NGC~6265 and NGC~6269 is quite high
($\sim700$ km s$^{-1}$), therefore NGC~6265 is almost certainly not a resident
of the group core. In any case NGC~6269 clearly dominates 
the optical light in AWM~5. Although it does not meet the formal definition of
a fossil group given in Jones et al. (2003), AWM~5 is a system with a dominant 
central galaxy, making it a group well suitable for a comparison with
both ``fossil'' and normal groups.

In a hierarchical Universe, smaller systems collapse and virialize 
at an earlier stage than more massive systems. Although smaller density fluctuations
also can generate groups of galaxies at later epochs, dense groups
should be generally older, more dynamically evolved systems than clusters 
of galaxies.
In galaxy groups, galaxy mergers can occur very efficiently because 
of their low velocity dispersions, comparable in some cases to the 
internal velocity dispersion of the individual member galaxies.  
In an old, relatively isolated group with little
subsequent infall, there may be sufficient time for most massive
galaxies (but not all the low mass galaxies) to lose energy via dynamical
friction and merge, producing a system consisting of a giant central
elliptical galaxy, dwarf galaxies and extended X-ray/dark matter halos.
This is one of the proposed scenarios for the formation of giant
isolated elliptical galaxies and it is based on the galactic
cannibalism discussed by Hausman \& Ostriker (1978).  
The observation of ``fossil'' groups (Ponman 
et al. 1994)
provides clear evidence for the above
mechanism. Numerical simulations (Barnes 1989) suggested that a few
billion years are required for compact group members to merge and form
a single elliptical galaxy. 

Presently, about a dozen such systems are identified (e.g. Ponman et al. 1994,
Vikhlinin et al. 1999, Mulchaey \& Zabludoff 1999, Jones et al. 2000, 
Jones et al. 2003) but just a
few have been studied in detail (e.g. Khosroshahi et al. 2004, Sun et al. 2004). 
In a 160 deg$^2$ ROSAT PSPC survey, Vikhlinin et al. (1999) found four such objects.
Computing the volume covered by their survey, they estimated that ``fossil'' groups
represent $\sim20\%$ of all clusters and groups with $L_X>2\times10^{43}$~ergs s$^{-1}$.
Consistent results were obtained
by Jones et al. (2003) who, using a precise definition of such a system, 
found that ``fossil'' groups represent a sizeable percentage (8\%-20\%) of
all systems with comparable X-ray luminosities, and are as numerous as
poor and rich clusters combined. \\
More recently Khosroshahi et al. (2007), using Chandra X-ray observations 
and optical data of a flux-limited sample of five fossil groups (plus 
additional systems from the literature), studied in detail the
scaling properties of fossils compared to normal groups and 
clusters. 
Considering groups having the same optical luminosity, they found that fossils 
are generally more X-ray luminous than non-fossil groups. In their sample,
however, fossil groups follow the 
conventional $L_X$-$T_X$ relation of galaxy groups and clusters, suggesting
that the 
X-ray luminosity and the gas temperature are both increased, as a consequence 
of their early formation. 
They also found other signatures of their early epoch of formation such as  
the higher mass concentration in fossils than in non-fossils,
and the $M_X$-$T_X$ relation, suggesting that fossils are hotter for a given total 
gravitational mass. 
The explanation they give for the difference between fossil and non-fossil properties
is that the cuspy potential well in fossils tends to raise the luminosity and 
temperature of their inter-galactic medium (IGM). However, this works together with the 
lower gas entropy (especially in lower mass systems) compared to normal groups, 
which could arise from less effective 
pre-heating of the gas in fossils. Another likely effect of the early shock heating
of the IGM is that galaxy formation efficiency is quite low in fossil systems,
as suggested by the high mass-to-light ratios observed.

In this paper, we describe the results from a Chandra observation of 
the galaxy group AWM~5, dominated by the central galaxy NGC~6269. 
The paper is organized as follows. We first describe the Chandra data preparation
and the analysis method in \S2. We present the results coming from the X-ray
morphology and spectral analysis of the diffuse gas in \S3, 
comparing the optical-IR properties of the galaxies in
the group with their X-ray properties in \S4. We discuss the results and their implications
in \S5. The ``stripped'' galaxy NGC~6265 falling into the
group is described in \S6, and the conclusions of the paper are drawn in \S7.

All the uncertainties are 1$\sigma$ (68\%) for one
interesting parameter, unless otherwise stated. 
The abundance estimates are relative to the
cosmic values given in Anders \& Grevesse (1989). 
Throughout this paper we assume $H_0=100\:h$ km s$^{-1}$ Mpc$^{-1}$,
$h=0.7$, $\Omega_m=0.3$ and $\Omega_\Lambda=0.7$.
For these parameters and a redshift of AWM~5 of $z=0.0348$, 
$1^{\prime\prime}\approx0.67$~kpc.


\section{Chandra Data Preparation and Analysis} \label{dataprep}


AWM~5 was observed by the ACIS-I detector on-board Chandra on 2003, December 29 for a total of 40~ks 
(ObsID: 4972).
The Chandra data analysis has been performed using CIAO 
v3.4, which applies the newest ACIS gain maps, the time-dependent ACIS gain correction, 
and the ACIS charge transfer inefficiency correction. 
The background light curve during the observation was examined to 
detect periods of high background 
following Markevitch et al. (2003a); no significant
flaring episodes were detected, giving a total good integration time
of $\sim39.5$~ks.

\subsection{Background Subtraction}\label{bkgsub}
Since the AWM~5 group fills most of the Chandra field of view, we use the 
blank-field observations, processed identically to the group observation 
(i.e. as described above) and reprojected onto the sky using
the aspect information from the group pointing. 
We renormalized the blank-fields 
to the background of the observation, using the ACIS-S2 chip, 
in a region of the ACIS field 
of view practically free from group emission. 
To perform the normalization, we used the energy band from 9.5 to 12 keV, which is
dominated by charged particles.

We followed the procedure of Vikhlinin et al. (2005), to check
if the diffuse soft X-ray background could be an important background component
in our observation. We extracted a 
spectrum from the source-free regions of the detector (again the S2 chip), 
subtracted the
renormalized blank-field background and fit the background subtracted spectrum in
XSPEC v11.3.2p (in the 0.4-1 keV band) with an unabsorbed {\tt mekal}
model, whose normalization was allowed to be negative.
The residuals were found to be negligible and consistent with zero, suggesting that
an additional soft background correction is not necessary for our dataset.

\section{Results}

\subsection{X-ray morphology and surface brightness}\label{radanal}
\begin{figure}
\plotone{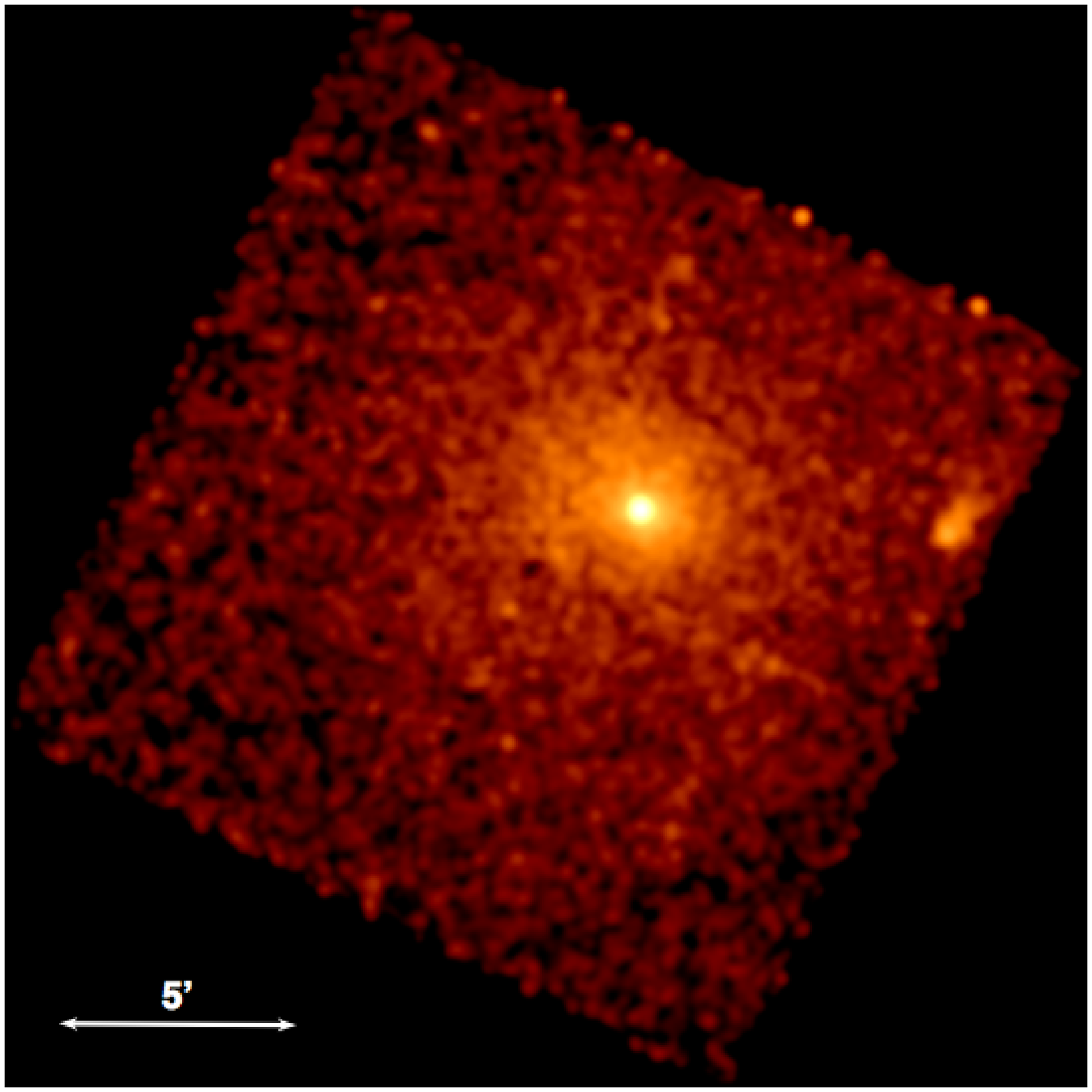}
\caption{The 0.5-4 keV Chandra ACIS-I image of AWM~5, with the X-ray peak centered
on the galaxy NGC~6269. The point sources were removed from the
exposure corrected image
and then smoothed with a 7 pixel ($\sim3.5^{\prime\prime}$) wide gaussian kernel. 
The group member NGC~6265 is
visible $\sim6^\prime$ to the west of NGC~6269, with a small ram-pressure stripped
tail extending to the north-west.\label{xrayimage}}
\end{figure}

\begin{figure}
\plotone{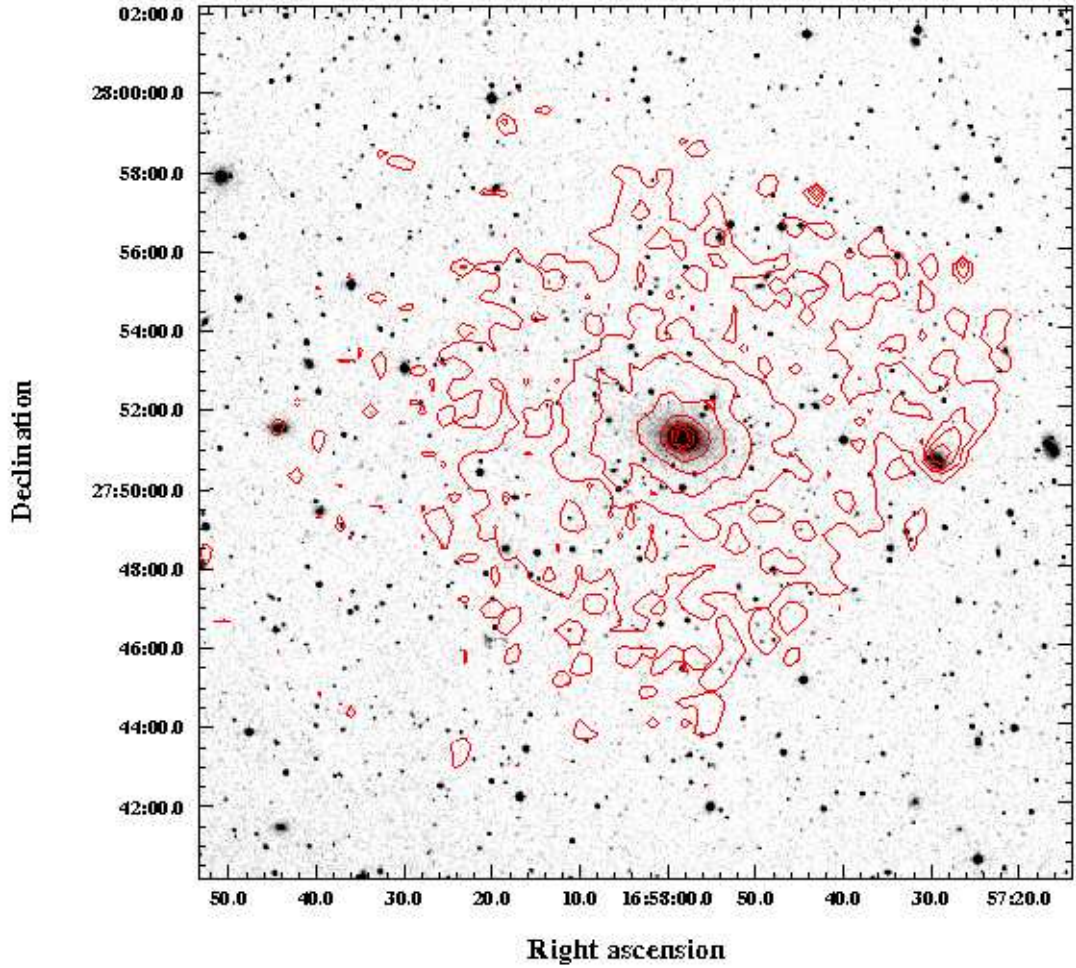}
\caption{Optical DSS image of AWM5 with the 0.5-4 keV Chandra ACIS-I contours
plotted in red. The central cD galaxy NGC~6269 is located in the peak of the X-ray
emission. The group member NGC~6265 is the only other galaxy in the group 
associated with extended X-ray emission.\label{opticalxray}}
\end{figure}

Figure~\ref{xrayimage} shows a 0.5-4 keV ACIS-I image of the center of AWM~5. The
X-ray peak is
coincident with the cD galaxy NGC~6269 (Fig.~\ref{opticalxray}). 
The group core is clearly visible at the center
of the image, and diffuse emission is visible out to several arcminutes from
the center. The X-ray tailed galaxy NGC~6265 (a member of the AWM~5 group) 
is visible on the far west of the image and is the only other group member associated
with extended X-ray emission.
We performed a radial analysis of the diffuse X-ray emission to determine its properties. 
We used CIAO $wavedetect$
to identify all the 3$\sigma$ point sources and excluded them from the image. To extract
a radial profile we used circular annuli of increasing thickness (from 
$4^{\prime\prime}$ to  $20^{\prime\prime}$), out to 
$\sim10^{\prime}$. 
We also computed an exposure map of the observation. 
The exposure map values were averaged inside each annulus
and renormalized to the exposure time in the central annulus.

\begin{figure}
\plotone{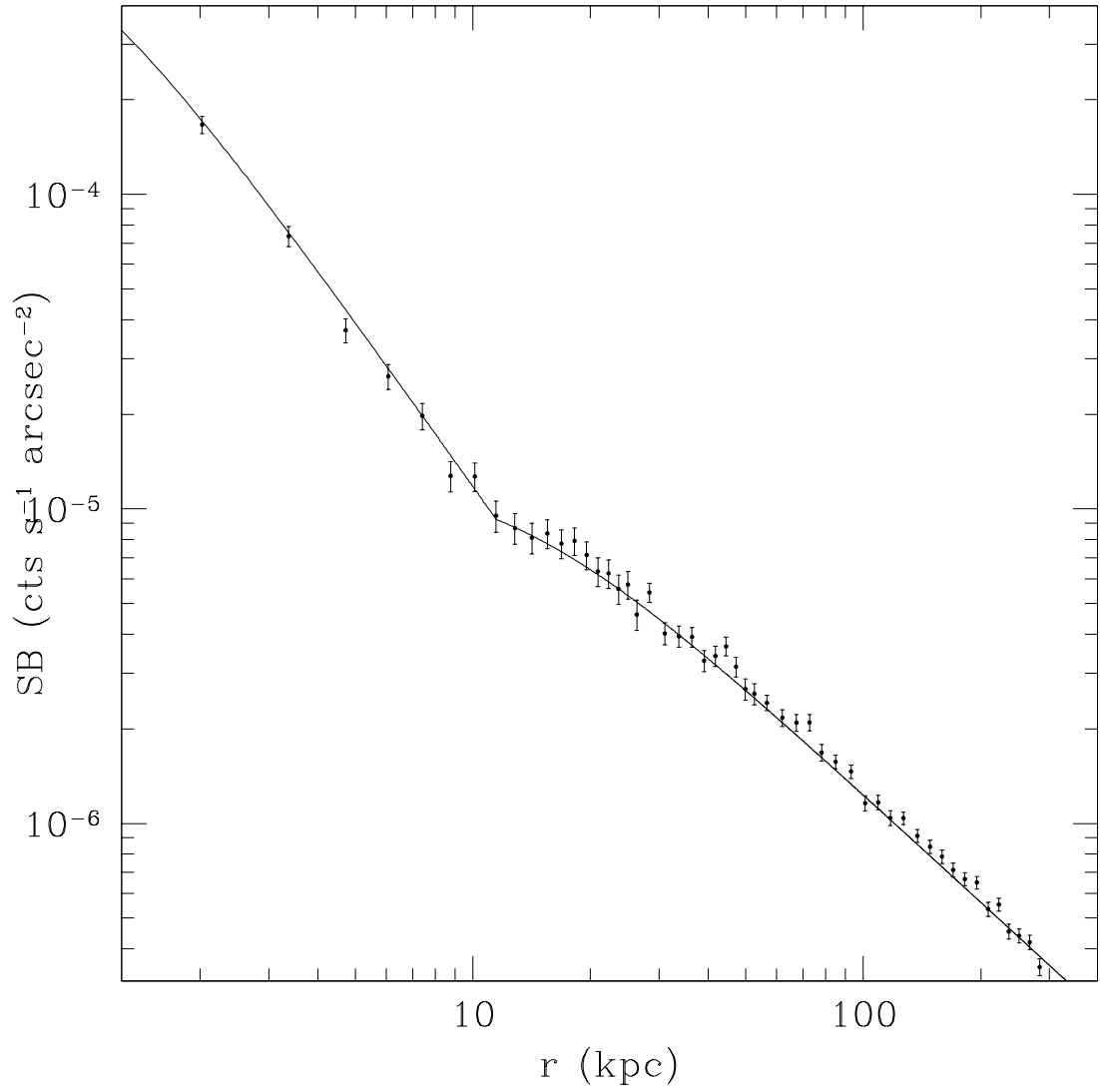}
\caption{0.5-4 keV Chandra ACIS-I surface brightness profile of the X-ray diffuse emission
from AWM~5. The profile shows a break in the slope at $\sim10$ kpc from
the center. The best fit double $\beta$ model (plotted as a solid line) has 
$r_c=0.8\pm0.1$ kpc, $\beta=0.458_{-0.013}^{+0.014}$ 
at $r<11.4$ kpc, and $r_c=12.8_{-0.9}^{+3.4}$ kpc, $\beta=0.357\pm0.002$ at $r>11.4$ 
kpc.\label{sbprofile}}
\end{figure}

The background subtracted exposure corrected 
surface brightness profile of AWM~5 is plotted in Figure~\ref{sbprofile}.
A break in the slope of the surface brightness profile is visible in the plot, 
at $\sim10$ kpc from
the group center. The surface brightness is well fit at all radii by two $\beta$ models
having core radii of $r_c=0.8\pm0.1$ kpc and $r_c=12.8_{-0.9}^{+3.4}$ kpc in the inner
($r<11.4$ kpc) and the outer part ( $r>11.4$ kpc) respectively ($\chi^2/dof=39.8/44$).
The corresponding best fit values of $\beta$ are $0.458_{-0.013}^{+0.014}$ and $0.357\pm0.002$,
respectively.

\begin{figure}
\plotone{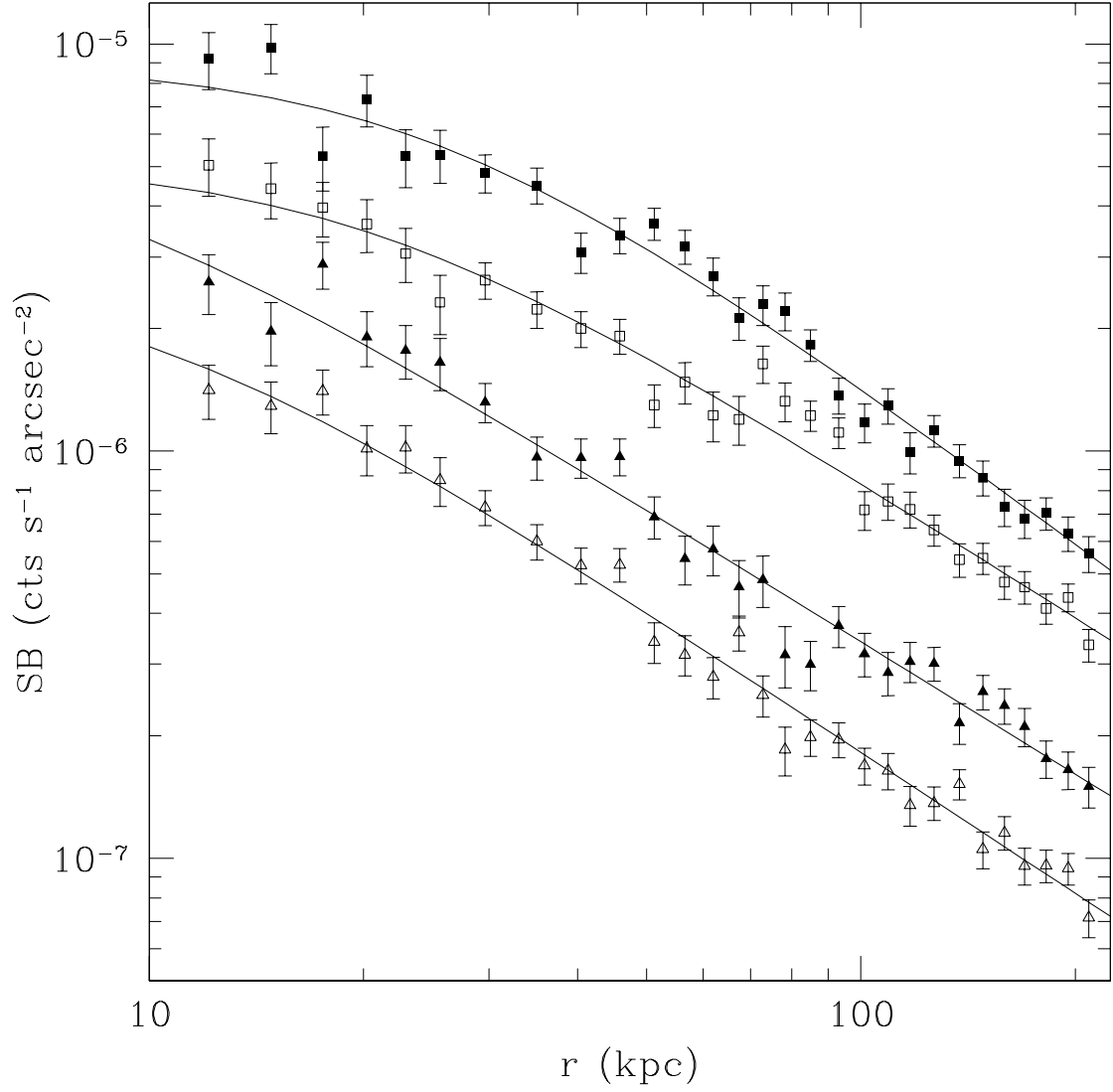}
\caption{0.5-4 keV Chandra ACIS-I surface brightness profile of the diffuse X-ray emission
from AWM~5 in four azimuthal sectors: north=filled squares, east=open squares,
south=filled triangles, west=open triangles. The eastern, southern
and western profiles are offset from their original value for clarity. The corresponding
best-fit $\beta$ model is shown in the plot as well.\label{figwedge}}
\end{figure}

To check for anisotropies in the X-ray emission we extracted radial profiles in four
azimuthal sectors (see Fig.~\ref{figwedge}). The procedure adopted to compute the surface 
brightness was the same
described above for the overall radial profile.
We fit a $\beta$ model to each sector at $r>10$~kpc (see Table~\ref{wedges}). 
While the northern and the western sectors
are clearly well described by single $\beta$ models, the southern and the eastern surface
brightness profiles are slightly more deviant. However the deviations are statistically
insignificant 
($\chi^2/\nu=34.2/26$ and $\chi^2/\nu=37.7/26$, respectively). 

\subsection{Mean Temperature and Virial Radius}\label{global}
The virial radius $r_{180}$ of AWM~5 can be estimated from:
\begin{equation}
\label{eqn1}
r_{180}=1.95h^{-1}\:(\langle kT \rangle/10\:keV)^{1/2}(1+z)^{-3/2} \:Mpc
\end{equation}
(Evrard et al. 1996)
This scaling relation may not be valid for cool groups (see e.g. Sanderson et al. 2003) but
can still be used for comparison with other work 
(see \S~\ref{comparegroups} below).
To compute the global temperature
$\langle kT \rangle$ necessary to estimate $r_{180}$, we extracted a spectrum
from 0.05$r_{180}$ to 0.3$r_{180}$.  
The central region is excluded
to avoid contamination from the group cooling
core (CC).
The spectra were fitted in XSPEC v11.3.2p with a 
single temperature {\tt mekal} model, with
Galactic absorption ($4.8\times10^{20}$ cm$^{-2}$; Stark et al. 1992) leaving the
metal abundance free to vary.
The values of $\langle kT \rangle$ and $r_{180}$ are
evaluated iteratively until convergence to a stable value of the
temperature is obtained ($\Delta kT\leq0.01$ keV between two different
iterations).
The best fit temperature was $kT=2.33_{-0.19}^{+0.39}$ keV, 
while the best fit metallicity was $Z=0.40_{-0.13}^{+0.22}$ $Z_\odot$.
The virial radius estimated from eq.~\ref{eqn1} for AWM~5 is $r_{180}=1.28$~Mpc.
The total X-ray luminosity observed within the emission region ($r<400$ kpc) is 
$\sim1.3\times10^{43}$ erg s$^{-1}$.
At $r_{500}$ (1.03 Mpc) and at $r_{180}$ 
the total luminosities, extrapolated from the surface brightness
$\beta$ model fit (\S~\ref{radanal}), are $\sim1.6\times10^{43}$ erg s$^{-1}$ and 
$\sim1.7\times10^{43}$ erg s$^{-1}$, respectively.

\subsection{Temperature and Abundance profiles}\label{ktzprof}
To study the physical properties of the group emission, we subdivided 
the emission from the group into annuli centered on the
X-ray peak.
We required $\sim1,000$ net counts in each of the central annuli
($r<40$ kpc), and $\sim2,000$ net counts in each of the outer annuli ($r>40$ kpc), 
where the surface brightness is lower and the contribution from the background is higher. 
We extracted a spectrum from each annulus (excluding
point sources) using
$specextract$, which
generates source and background spectra and builds appropriate RMFs and ARFs. The background
is taken from the re-normalized blank field observations using the same region as the 
source.

The spectra were analyzed with XSPEC v11.3.2p (Arnaud et al. 1996) and fitted with a 
single-temperature {\tt mekal} model (Kaastra 1992; Liedahl et al. 1995).
The free parameters in the model are 
the gas temperature $kT$, the gas metallicity $Z$ and the normalization. 
The spectral band considered is 0.5-4 keV. 
All the spectra were rebinned to have at least 20 counts per bin.
We have measured $N_H$ from the X-ray data, finding it consistent (within 1$\sigma$) 
with the Galactic value
along the line of sight, as derived from radio data 
($4.8\times10^{20}$ cm$^{-2}$; Stark et al. 1992).
Since the presence of AGN activity in the center of AWM~5 is very likely, we have tried
to add an additional power-law component to the thermal model in the
central annulus. The fit is not improved significantly by the presence of 
the additional spectral component ($\Delta \chi_{\nu}^2 < 0.05$) and the 
contribution of a power-law is $<10$\% of the thermal component.
We also considered an image in the 3-8 keV band, finding that no hard
point-like emission is present both in the Chandra 40ks observation and in an 
off-axis XMM archival observation (centered on NGC~6264, PI: Greenhill).

\begin{figure}
\plottwo{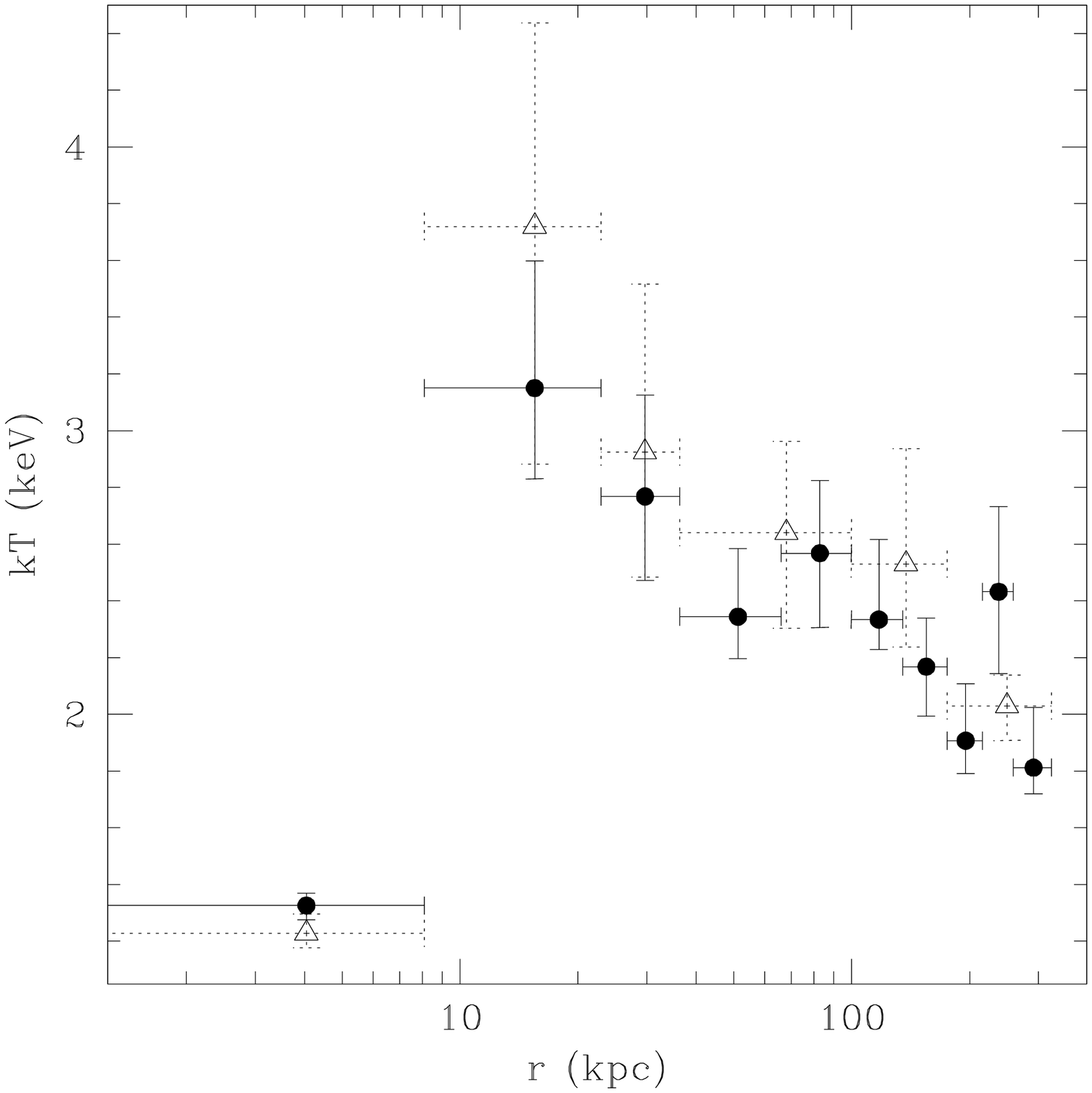}{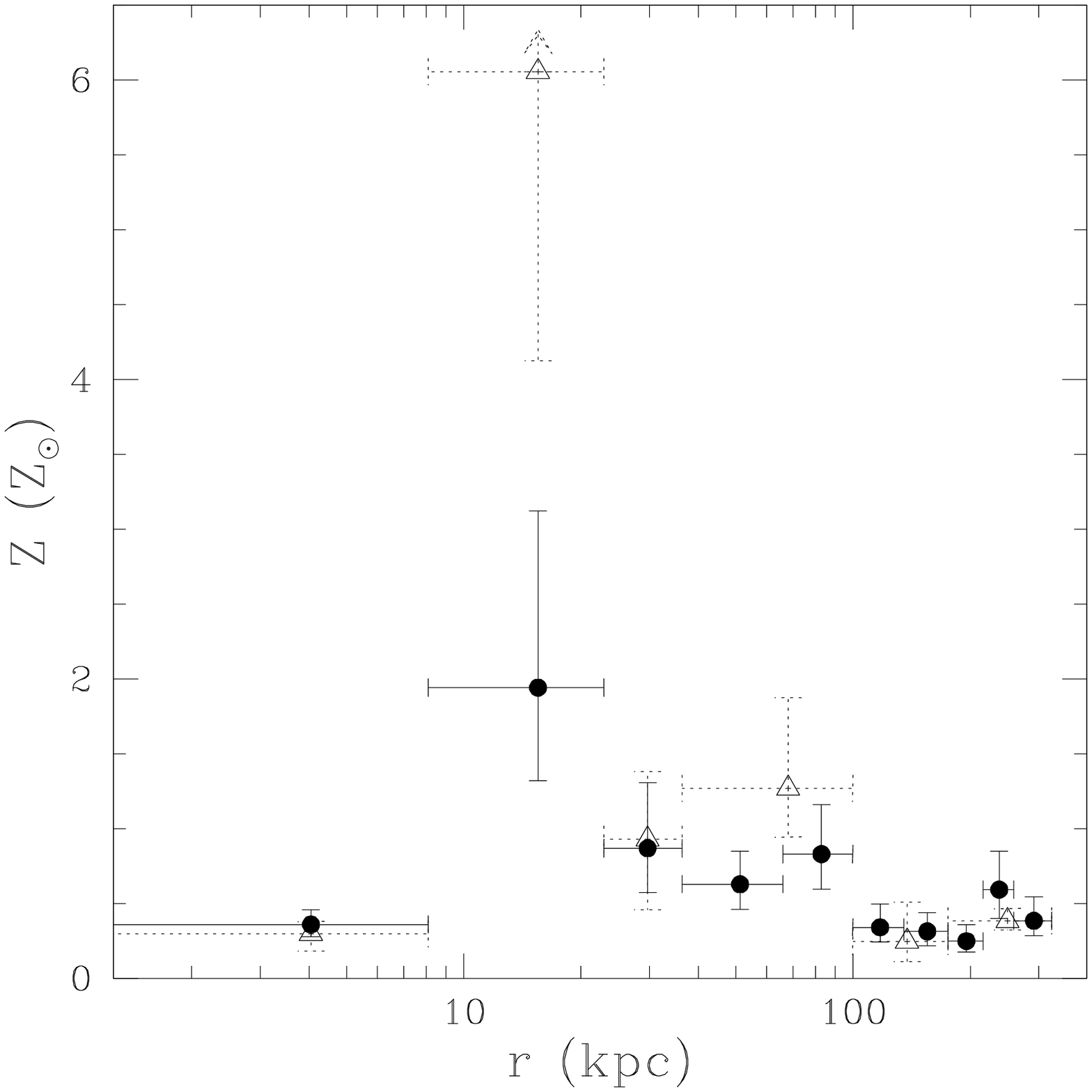}
\caption{{\it Left:} Temperature profile for the X-ray gas observed in AWM~5. The projected
temperature is plotted with filled circles and solid error bars, while the corresponding
deprojected temperatures are plotted with open triangles and dotted error bars. A cooling
core $\sim8$ kpc in size is visible in both the projected and deprojected
profiles, with the latter showing a stronger temperature gradient toward the center. The two
profiles are in general agreement at all radii and show a decline of the temperature from $\sim3.5$~keV,
just outside the core, to $\sim2$~keV, at $\sim300$~kpc from the center. 
{\it Right:} Abundance profile
for AWM~5. The symbols have the same meaning as in the left panel. Both the projected and deprojected
profiles suggest an abundance ``hole'' interior to $\sim10$~kpc, where the temperature declines
sharply.\label{ktz}}
\end{figure}

The projected temperature and abundance profiles of AWM~5 
are shown in Figure~\ref{ktz}.

The temperature profile shows the presence of an $\sim$8~kpc-sized CC (projected $kT\sim1.3$
keV; de-projected $kT\sim1.2$) with the profile having a peak just outside the core and then declining from
$\sim3$~keV to $\sim2$~keV at larger radii.
The abundance profile is suggestive of a ``hole'' in the center coincident with the CC. The
abundance value rises just outside the core reaching a maximum of a few times solar
at a radius of 
$\sim15-20$~kpc and then declines to subsolar values. The probability that the value of the abundance
is the same in the inner two bins and that the ``hole'' is produced by statistical fluctuations
is $\sim1\%$ and $\sim0.3\%$ in the projected and deprojected case, respectively. 
Similar behaviour is observed in other groups and clusters of galaxies
(e.g. Perseus cluster; Schmidt et al. 2002; Churazov et al. 2001; 2002) and
was observed in ``fossil'' groups (e.g.
ESO 3060170; Sun et al. 2004).

We also examined the radial variation of Si, S and Mg fitting a {\tt vmekal} single
temperature model to the radial annuli. The Fe and Mg abundance show
a trend very similar to what
was observed in the metal abundance profile in the {\tt mekal} fit. On the other hand
S and Si abundances are constant within the errors, although the
large statistical errors cannot exclude that these abundances could 
follow the same radial trend as the
other metals. The data quality does not allow a direct comparison 
of the SNIa and SNII product yields with the
results coming from the study of Rasmussen \& Ponman (2007) 
on a large sample of galaxy groups.

\subsection{Physical Parameters}\label{physpar}

The electron density profile can be obtained by deprojecting the surface
brightness profile (see e.g. David et al. 2001; Sun et al. 2003).
Based on the ROSAT All-Sky Survey, we determined that the flux contribution 
from regions beyond the outermost bin ($\sim10^\prime$) is not significant.
Thus the ``onion-peeling'' technique adopted starts in the outer annulus, converting
the observed background-subtracted and exposure-corrected 0.5-4 keV surface brightness 
profile (derived in 22 radial bins) to electron density. 
In each bin, an emissivity for the gas at a fixed temperature $kT$ 
and $Z$ is computed for a standard {\tt mekal} model to perform the conversion. 
The procedure determines the density at progressively smaller radii, 
subtracting the projected emission from larger radii.
In the temperature range observed in the projected profile of AWM~5 ($1.3\leq kT\leq3.2$ keV), 
the X-ray emissivity
is sensitive to both abundance and temperature. 
Therefore, with the assumption of spherical symmetry, we first performed a spectral 
deprojection to obtain the 3D temperature and abundance profiles (Fig.~\ref{ktz}). 
Only six radial bins were considered for the spectral deprojection. 
Because of the steep temperature gradient observed in the core of AWM~5 we chose a broken
power-law model to fit $kT$. A second order polynomial was chosen to model the radial
dependence of the abundance. We varied the values of $kT$ and $Z$ within their $1\sigma$ 
uncertainties,
through 1000 Monte Carlo simulations to obtain the $1\sigma$ errors on the best-fit
parameters of the fits to the $kT$ and $Z$ profiles. 
The statistical errors in the electron density 
were determined by combining the uncertainties in the X-ray 
emissivity and in the surface brightness profiles.
The missing emission volume
in each annulus because of the chip edges, gaps and point sources was derived
by Monte Carlo simulations and included in the computation of the errors as well.

\begin{figure}
\plotone{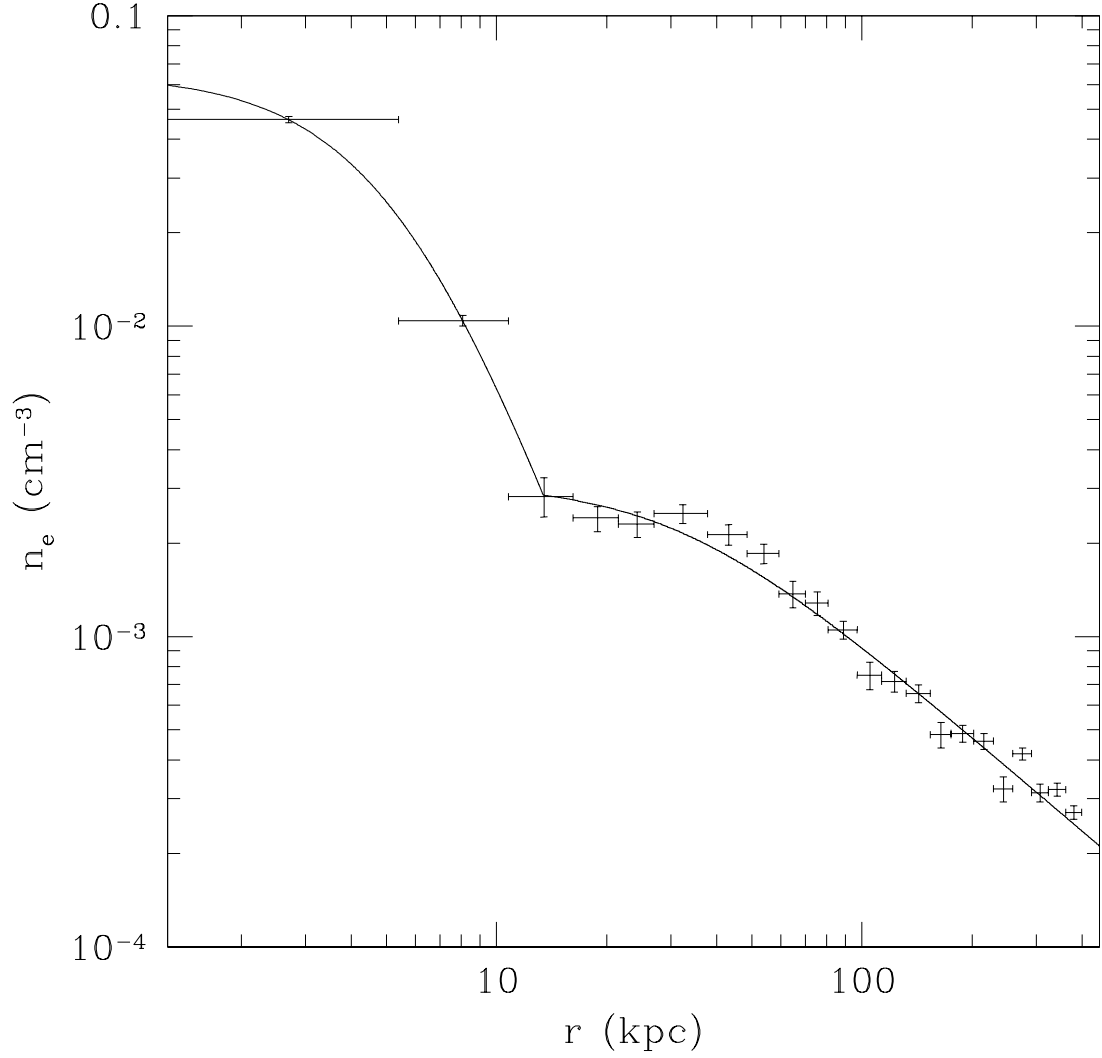}
\caption{Electron density profile derived from a deprojection of the surface brightness.
The density profile shows a clear break in the slope at $r\sim10$~kpc and can be fit
by a double $\beta$-model. At $r<13.4$~kpc we obtain
$\beta=0.72_{-0.11}^{+0.16}$ and $r_c=5.7_{-1.5}^{+1.8}$~kpc, while at $r>13.4$~kpc
the best fit value of $r_c$ is $31.3_{-5.5}^{+5.8}$ kpc and $\beta=0.34\pm0.01$.
We found that $n_{\rm e}\propto r^{-0.91 \pm 0.02}$ between 30 kpc and 300 kpc.\label{ne_prof}}
\end{figure}

\begin{figure}
\epsscale{.60}
\plotone{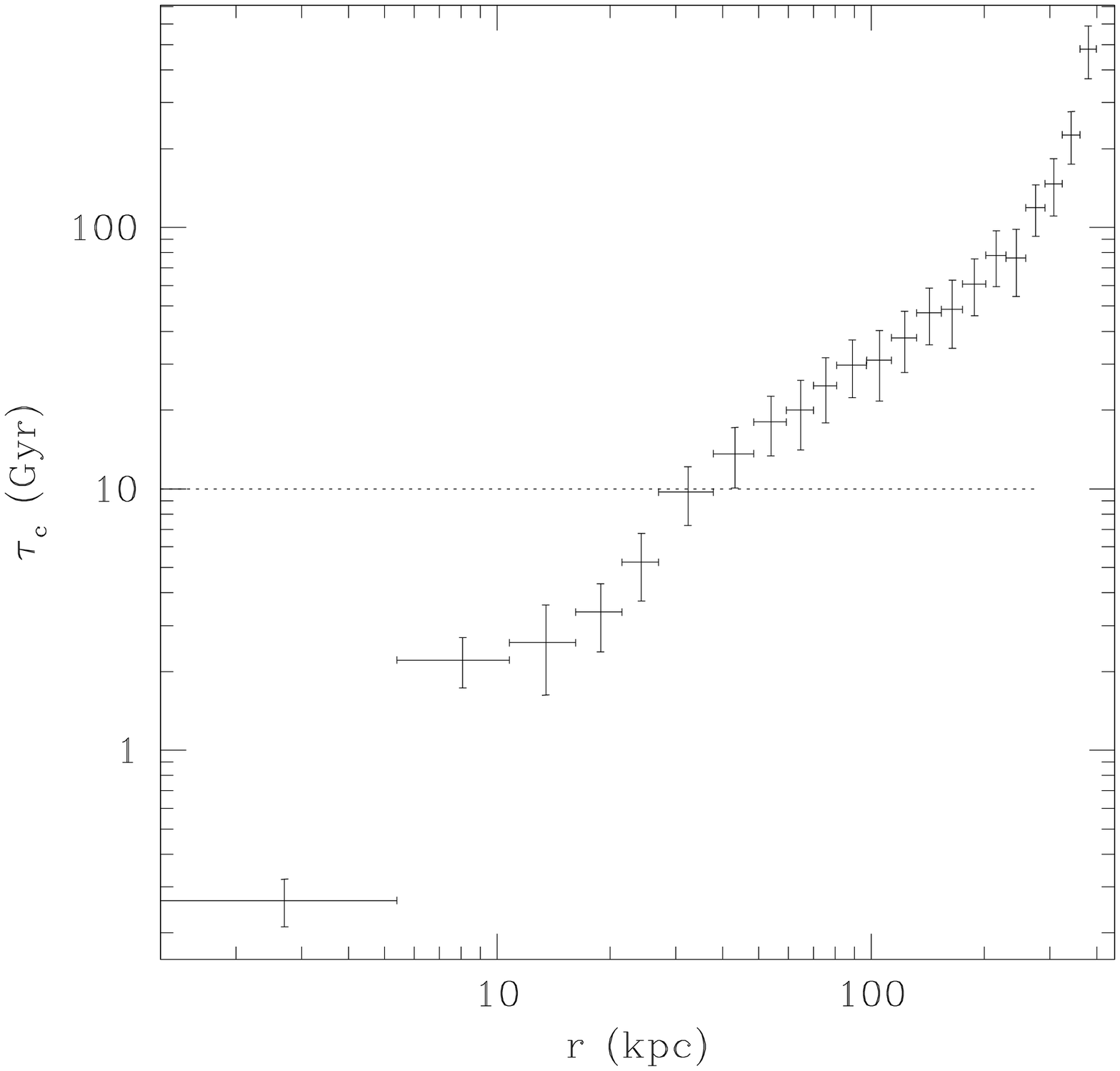}
\plotone{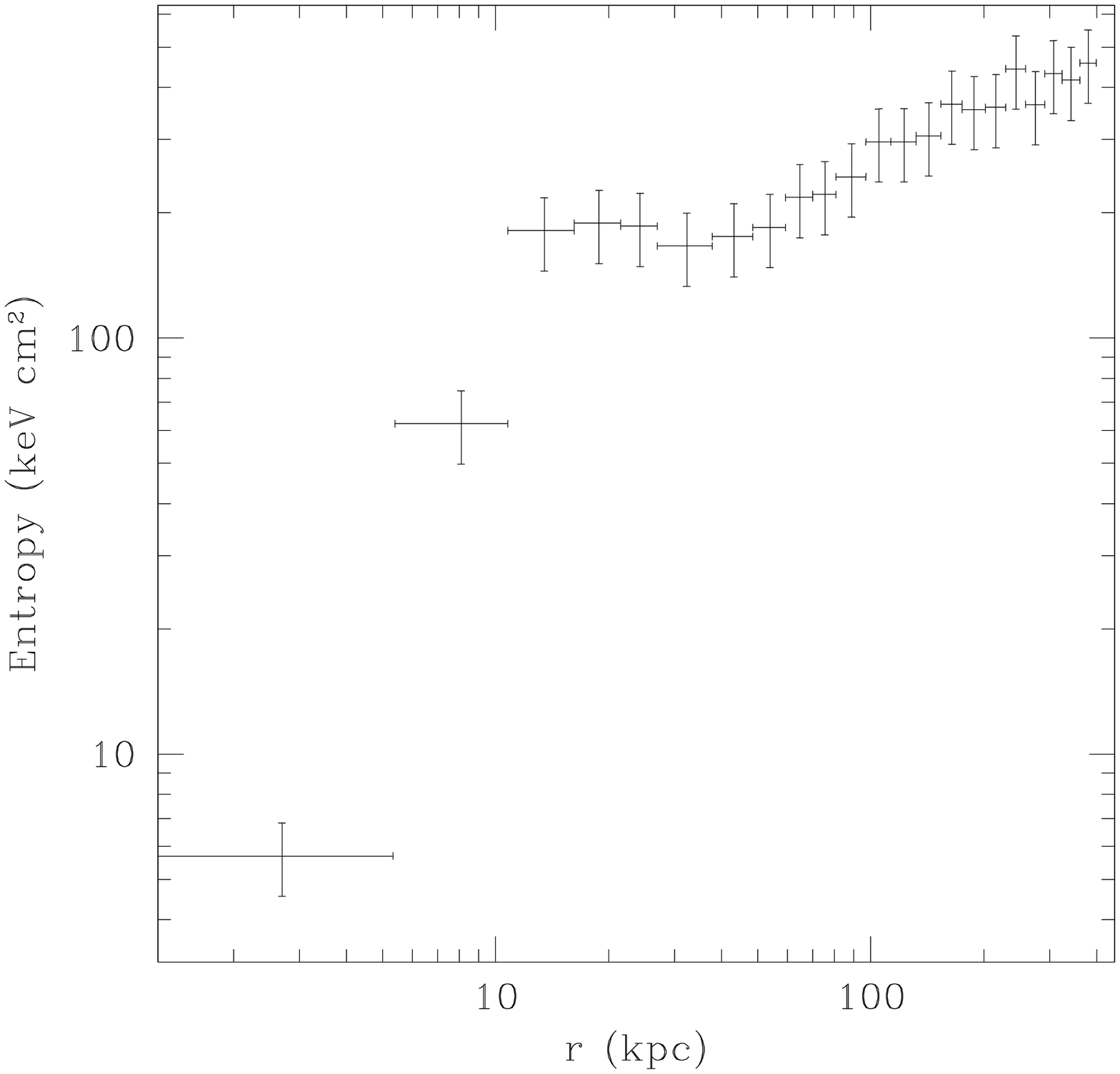}
\caption{{\it Top:} Cooling time $\tau_c$ of the gas as a function of the radial distance. 
The cooling time is less than a Hubble time inside the central $\sim$30~kpc. A comparison with the
size of the cool core observed in the temperature profile ($\sim8$~kpc) may indicate that
the central regions of AWM~5 were re-heated.
{\it Bottom:} Entropy of the gas as a function of the distance from the center of the group.
A flattening of the entropy profile is observed just outside the center and is just partly 
due to the cavities observed in correspondance with the radio lobes, and most likely related to the
re-heating of the core.
\label{tc_entr}}
\end{figure}

The derived electron density profile is shown in Fig.~\ref{ne_prof}. It can be fit by
a double $\beta$-model although the best-fit $\chi^{2}$ is not good ($\chi^{2}/dof$ = 47.0/16),
mostly because of the three points at 25~kpc$<r<$60~kpc. The inner $\beta$-model (at $r<13.4$~kpc)
has best fit values of: 
$\beta=0.72_{-0.11}^{+0.16}$; $r_c=5.7_{-1.5}^{+1.8}$~kpc). 
A core radius of $r_c=31.3_{-5.5}^{+5.8}$ kpc and a $\beta=0.34\pm0.01$ were derived 
for the outer $\beta$-model (at $r>13.4$~kpc).
At $r>30$~kpc a power-law can also be fit to the data ($\chi^{2}/dof$ = 33.3/15).
We found that n$_{\rm e} \propto$ r$^{-0.91 \pm 0.02}$ between 30 kpc and 300 kpc.
The slope 
measured for the density profile
is very similar to that observed in cool groups
(e.g. NGC1550, Sun et al. 2003) but much flatter than those measured in ``fossil'' groups
(e.g. NGC6482, Khosroshahi et al. 2004; ESO3060170, Sun et al. 2004) and in clusters, where
generally n$_{\rm e} \propto$ r$^{-2}$ beyond several core radii. With the
deprojected electron density profile, we were also able to derive the
cooling time and entropy profiles (defined as $S = kT/n_{\rm e}^{2/3}$; Fig.~\ref{tc_entr}).
The cooling time is less than 10$^{9}$ yr in
the very center and less than a Hubble time ($\sim$ 10$^{10}$ yr) within
the central 30 kpc. 

\subsection{Energetics of the radio lobes}\label{radio}

\begin{figure}
\plotone{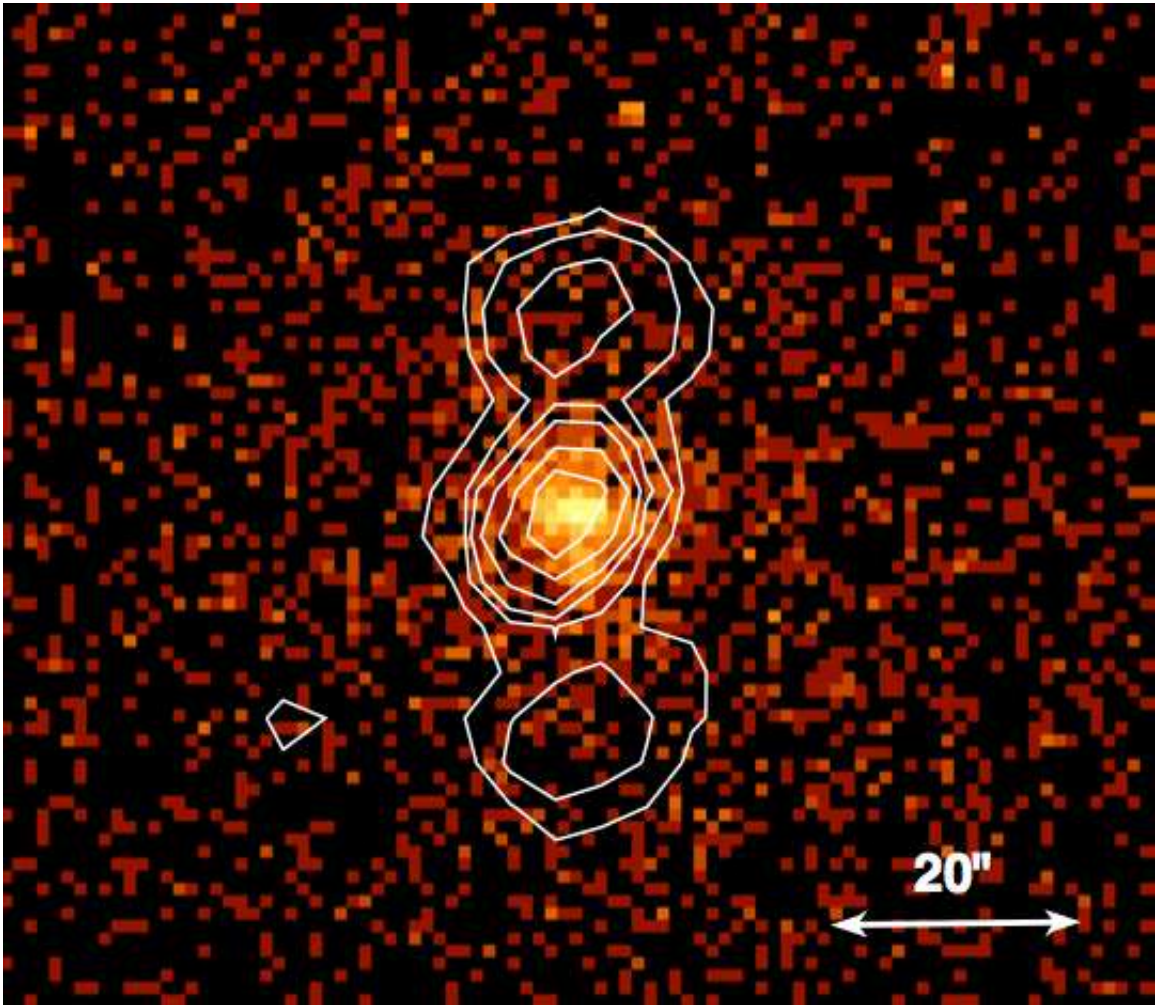}
\caption{1400~MHz radio contours from the FIRST superimposed on the ACIS-I 0.5-4 keV
image. The radio source has a central nucleus ($\sim27$ mJy) coincident with the group core,
with north-south radio lobes ($\sim8$ and $\sim10$ mJy) extending 
25$^{\prime\prime}$ ($\sim17$~kpc) from the center.\label{radioimg}}
\end{figure}

As described in \S~\ref{intro}, AWM~5 hosts a weak central radio source
(Fig.~\ref{radioimg}). The two lobes seem to be coincident with a depression in the
X-ray emission, observed especially in the southern lobe. However the Chandra data
are not sufficiently deep to perform any detailed analysis on such cavities.

From the geometrical properties of the radio lobes and the density and 
temperature (and hence pressure) profiles derived above we can determine 
some of their physical parameters.
The calculations derived in this section refer to the northern radio lobe,
however the two lobes are quite symmetric and we can consider these numbers
valid also for the southern lobe.
The northern lobe shape can be approximated with a sphere 
with a $\sim8^{\prime\prime}$ radius ($\sim5.4$~kpc). We assume that
the lobe rose buoyantly at half the speed of sound (e.g. as in M~87, Churazov et al. 2001).
At the temperature of the
CC of AWM~5, $kT=1.2$~keV, the speed of sound is $c_s\sim600$ km s$^{-1}$,
the rise time of the lobe and hence the age is therefore $3.5\times10^7$ yr, consistent
with the radiative age $t_{rad}=6.5\times10^7$ derived by Giacintucci et al. (2007).
The energy necessary to inflate the lobe can be calculated
as 4$pV$ ($p$ is the local hot gas pressure since the lobe is supposed to be in pressure
equilibrium, and $V$ is
the volume of the lobe itself) and is $6.4\times10^{57}$ erg. Therefore the
power necessary to inflate the lobe in $3.5\times10^7$ yr 
is $\sim5.8\times10^{42}$ erg s$^{-1}$. This number
is two orders of magnitude higher than the X-ray upper limit we obtain for 
the AGN at the center of NGC~6269. 
If we estimate the mechanical luminosity of the rising bubble as $L_{mech}=pV/t$ (B\^{\i}rzan
et al. 2004), where $t$ is the age of the bubble, we obtain 
$L_{mech}\sim1.4\times10^{42}$~erg s$^{-1}$. This can also be compared with a total radio luminosity
of $\sim10^{40}$~erg s$^{-1}$ (see \S~\ref{reheat}), in agreement with the $L_{radio}-L_{mech}$
relation derived by B\^{\i}rzan et al. (2004), who find that $L_{mech}\gg L_{radio}$.

Marconi \& Hunt (2003) derived a correlation 
between the bulge infrared luminosity ($L_{K_s,bulge}$)
and the mass of the central black hole ($M_{BH}$).
For NGC~6269 we estimate $M_{BH}\sim10^9$~M$_\odot$.
The luminosity expected for steady Bondi accretion (Bondi 1952; see
\S~\ref{bondi}) of
a $10^9$ solar masses black hole (deriving the density at $r=0$ from the
inner $\beta$-model obtained in \S~\ref{physpar}) is at least one order of magnitude larger than
the energy necessary to inflate the radio bubbles.

If the radio-lobes are not parallel to the plane of the sky the estimates performed above will
be different. In particular, if we name $\theta$ the inclination angle of the lobe axis
with respect to the plane of the sky, the volume of the bubble will be increased by a factor
$(cos\:\theta)^{-1}$ and so the energy necessary to inflate the bubble. However, unless we are dealing
with a very extreme inclination angle, the estimates computed in this Section can be 
considered reliable within a factor of a few.

\subsection{X-ray Gas Mass and Total Gravitating Mass}

\begin{figure}
\epsscale{.60}
\plotone{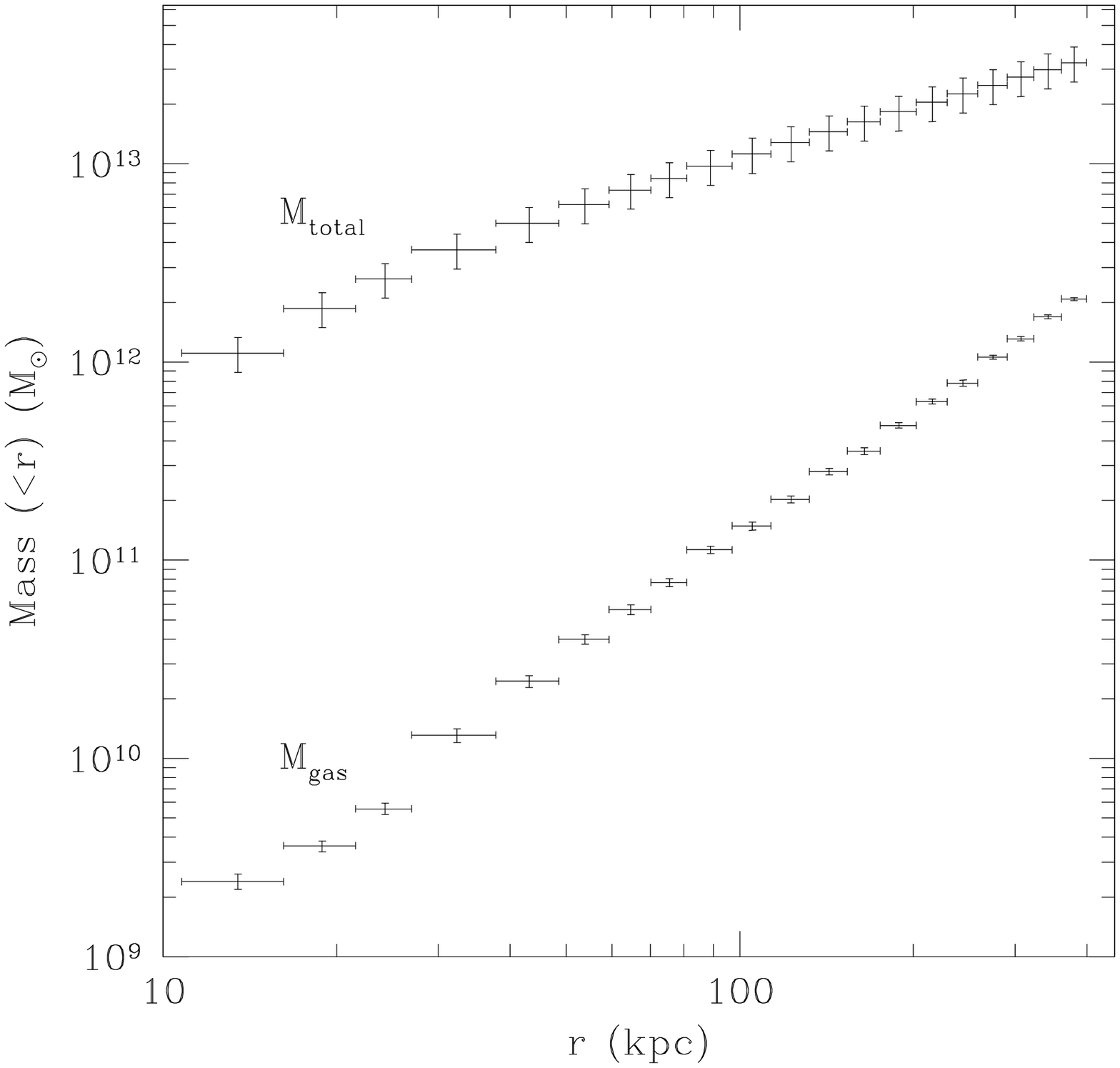}
\plotone{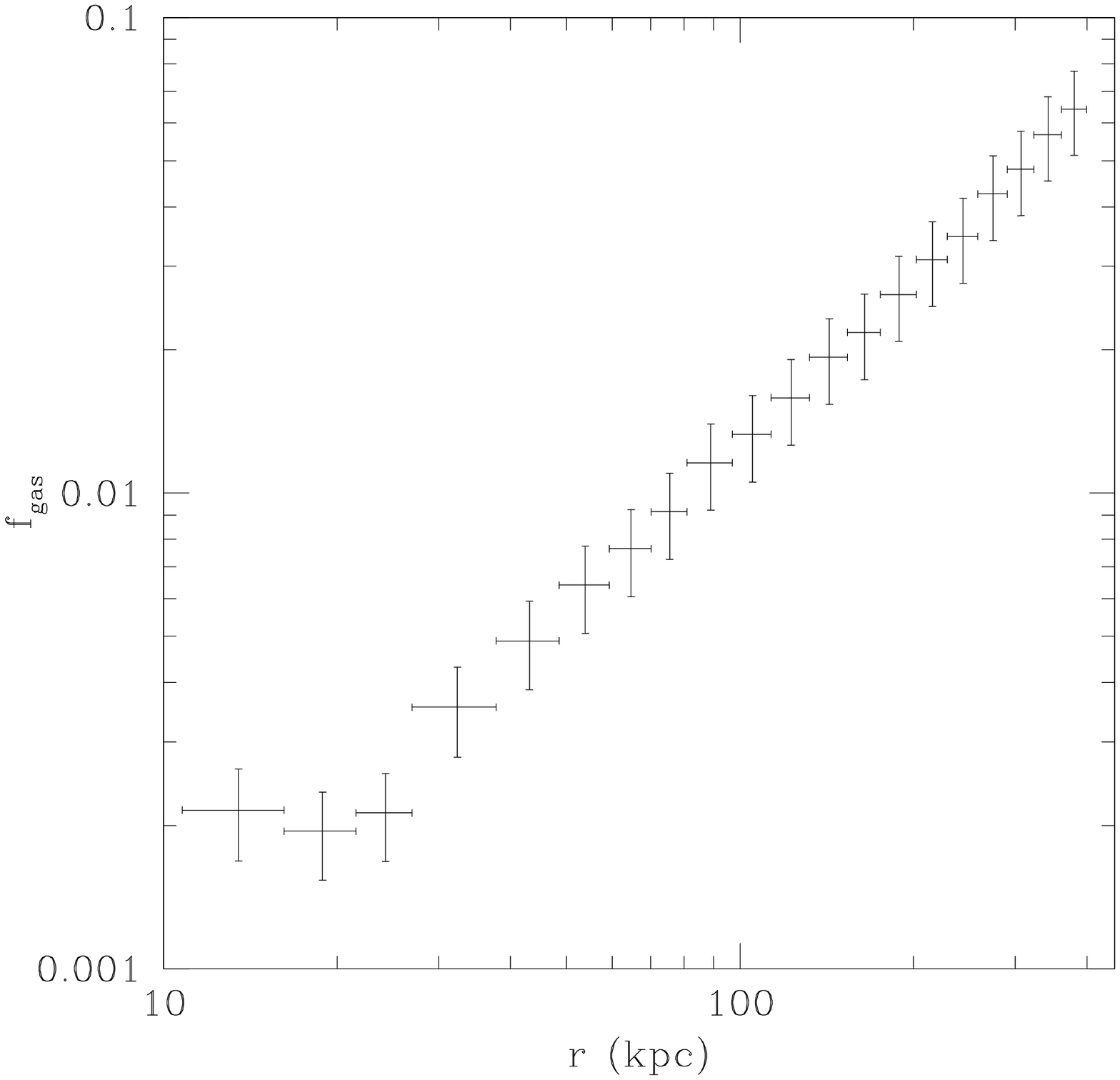}
\caption{{\it Top:} Total gravitating mass and X-ray gas mass enclosed within a radius $r$ in the
AWM~5 group.
{\it Bottom:} Fraction of the gas mass over the total gravitating mass as a function of the distance
from the center. The gas fraction is flat within the central 30~kpc and increases afterwards reaching
a maximum value of $6.5\%$ at $\sim380$~kpc.
\label{massprof}}
\end{figure}

The gas density and temperature profiles derived in \S~\ref{physpar} 
can be used to estimate the total
gravitational mass profile under the assumption of hydrostatic
equilibrium:
\begin{equation}  
M (<r) = - \frac{k T(r) r}{\mu m_p G} \left(\frac{d \log n_e(r)}{d \log r} + 
\frac{d \log T(r)}{d \log r}\right)
\end{equation}
where $G$ and $m_p$ are the gravitational constant and proton mass,
and the mean particle mass is $\mu=0.6$.
Following a procedure similar to that applied in \S~\ref{physpar} in the
deprojection of the surface brightness profile, we considered the broken power-law 
functional $kT(r)$ as derived in \S~\ref{physpar}. 
We perturbed the value of $kT(r)$ within its $1\sigma$ 
uncertainties, through 1000 Monte Carlo simulations. 
For the electron density profile we assumed the double
$\beta$-model described above in \S~\ref{physpar} and a similar
method was applied to derive the density gradient and its
uncertainties. The derived uncertainties were combined to compute
the errors for the total gravitating mass profile. 
Both the total mass and gas mass profiles are shown in Fig.~\ref{massprof}, along
with the gas mass fraction. 
In these plots, we did not show the CC region ($r<10$~kpc) in which the
determination of the mass is obviously more difficult because of the strong 
temperature gradient.

The total mass within the central ~15 kpc is quite high ($\sim10^{12}$ M$_\odot$),
a value higher than those usually observed in groups (see e.g. Gastaldello et al. 2007).
However AWM5 is a hot group and the high mass 
concentration in the center is most likely 
due to the presence of the central massive galaxy NGC6269, whose hot gas halo is 
superimposed over the group halo.
We fitted a Navarro, Frenk \& White (1996; NFW) profile to measure the concentration 
parameter $c_{200}$ as in Khosroshahi et al. (2007). A single NFW fit cannot account for 
the shape of the mass profile at all radii. This can be explained by superposition
of two halos of dark matter from the group and from the central galaxy NGC~6269.
Moreover complications could be introduced by the radio source (which could
be more extended than visible in the FIRST) and by the presence of a steep
gradient in temperature. Fitting the inner part ($r<80$ kpc) we obtain
an unusual concentration parameter $c_{200}\sim100$. On the other hand, if we
fit the outer part ($r>80$ kpc) we obtain a more reasonable $c_{200}\sim20$ which 
in a $M$ vs. $c_{200}$ diagram falls in the region occupied
by other groups (e.g. the RXJ0454.8-1806 group; Khosroshahi et al. 2007).

The gas mass fraction has a constant value ($\sim0.05\%$) at $10$~kpc$<r<30$~kpc, 
starting to increase rapidly at $r>30$~kpc. The gas fraction reaches a maximum
value of $\sim6.5\%$ at $\sim380$~kpc, with no obvious signs of flattening at large radii.
For comparison, typical gas fractions at large radii are $\sim15\%$ in clusters (e.g.
David et al. 1995) and $\sim5\%$ in groups (Gastaldello et al. 2007).
In the ``fossil'' group ESO3060170 (Sun et al. 2004) the gas fraction is $\sim5\%$ at 
$r=200$~kpc 
and stops growing after that radius. In AWM5, $f_{gas}$ is $\sim2.5\%$ at 
$r=200$~kpc, however it shows a trend to increase even 
at $r>380$~kpc and it might converge to a larger value than observed with the Chandra
data.

\subsection{Entropy profile}\label{comparegroups}

\begin{figure}
\epsscale{.60}
\plotone{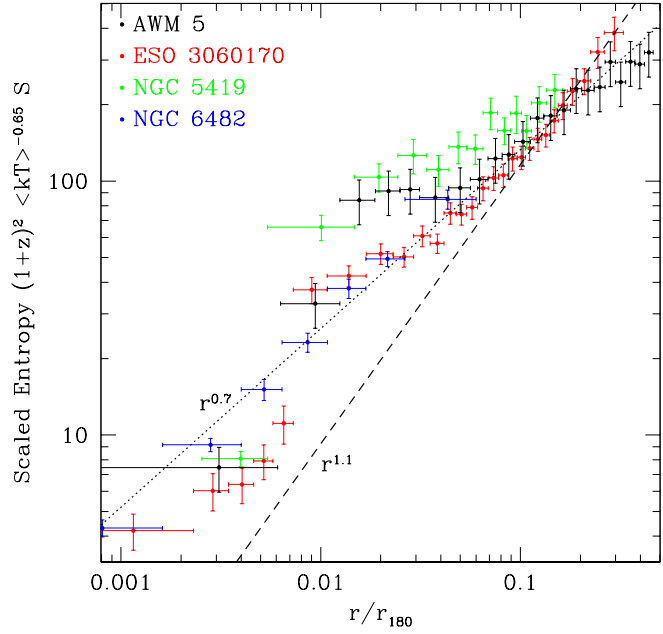}
\caption{Scaled entropy profiles for AWM~5 compared with other groups (ESO3060170,
NGC~5419, NGC~6482). The entropy profile of AWM~5 is generally flatter
than the other groups ($\propto r^{0.7}$, dotted line) and also from the slope
($\propto r^{1.1}$, plotted as a dashed line) predicted by cluster simulations (Tozzi \& Norman 2001). 
The flattening
of the entropy profile just outside the core of AWM~5 ($0.01r_{180}\leq r \leq 0.05r_{180}$)
has been observed also in other fossil groups
lacking a cool core (e.g. NGC~5419, ESO3060170) and is most likely related to the re-heating of the cores.
\label{scalentr}}
\end{figure}

The entropy profile of AWM~5 can be compared with those
of other groups. 
We plot the scaled
entropy profiles ((1+z)$^{2}T^{-0.65}S$) of AWM~5 and three other
groups in Fig.~\ref{scalentr}: ESO 3060170 (Sun et al. 2004) 
NGC~5419 and NGC~6482 (Khosroshahi et al. 2004). 
The scaled profile of AWM~5 is clearly flatter than all but NGC~5419. Simulations 
and analytical models of spherical accretion
onto clusters predict entropy profiles with a $S\propto r^{1.1}$ (Tozzi \& Norman 2001), 
at least outside
the central region (where heating and cooling are important). However, in the case of AWM~5, the
entropy profile is never consistent with such a slope. The entropy is better described
by a slope $S\propto r^{0.7}$. 

At 0.01 $r_{180}\: < r \:<$ 0.04 $r_{180}$ a further flattening of the profile is observed, similar
to what is observed in ESO 3060170 and NGC~5419. This variety
of inner entropy profiles is generally 
related to the evolutionary stages of group
CCs. Groups with heated gas cores can have flat inner entropy
profiles (see e.g. AWM 4, O'Sullivan et al. 2005), while those with large
CCs do not have flat entropy cores.
In the case of AWM~5 the flattening of the entropy, at least in the inner part, is coincident
with the position of the X-ray cavities
generated by the two radio-lobes. However the flattening extends far beyond those cavities 
and most likely is directly related to the re-heating
of the CC (see \S~\ref{reheat}).


\section{Optical-IR properties vs. X-ray properties}

\begin{figure}
\epsscale{0.95}
\plotone{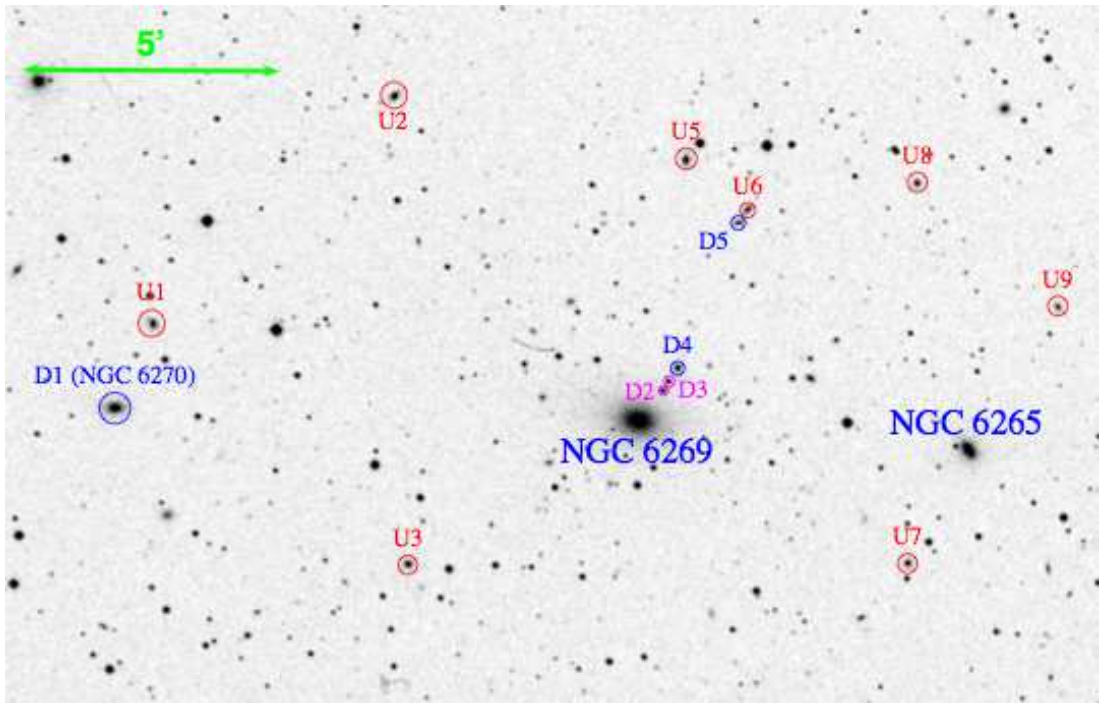}
\caption{ESO Digitized Sky Survey R-band image of the group AWM~5. The blue circles refer
to the group members (with a confirmed radial velocity from Koranyi \& Geller 2002) 
with a detected X-ray counterpart from Chandra ACIS-I, while the group members
without a Chandra counterpart are shown with red circles. The two galaxies
indicated with a magenta circle do not have a determination of the radial velocity,
however they are detected in the Chandra image and look to be part of an alignment together
with the source D4, pointing toward NGC~6269.
\label{optimage}}
\end{figure}

Figure~\ref{optimage} shows the ESO Digitized Sky Survey image (R band) of the central regions 
of AWM~5. Galaxies with measured radial velocities (from
Koranyi \& Geller 2002) consistent with that of the group are labelled. 
Apart from NGC~6269 (the dominant galaxy in AWM~5) and NGC~6265 (see \S~\ref{6265}),
only three of the 12 group galaxies in the Chandra field of view are detected in the 
X-rays. 
Two other X-ray sources within 1$^\prime$ ($\sim40$~kpc) of NGC~6269 have red, galaxy-like
optical counterparts and are likely group members.
These two objects together with the galaxy D4
are listed in Table~\ref{optical}. Their optical colors ($u-g=2.5$; 
$g-r=1.4$, $r-i=1.1$) rule out the possibility that they could be background AGN (see e.g. 
Richards et al. 2001). These values put them in the region of the color-color diagrams of
late M-type stars (Finlator et al. 2000), and are very different from the colors observed
in all the other galaxies of the group ($1.7<u-g<2.0$; $0.8<g-r<0.9$, $0.4<r-i<0.5$). 
The presence of a Chandra counterpart excludes the possibility that we are observing two red giants
and suggests that we are dealing with very old stellar systems.

\begin{figure}
\epsscale{.60}
\plotone{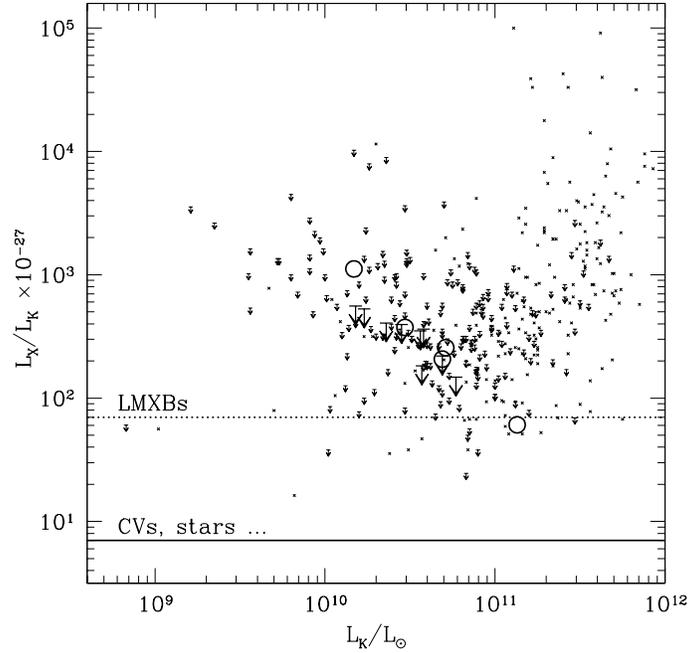}
\caption{Infrared luminosity ($L_{K_s}$, expressed in solar units) vs. the ratio of
the X-ray luminosity over $L_{K_s}$ 
in the AWM~5 members (big circles and
big bold upper limits) and in a sample of early-type galaxies from Ellis \& O'Sullivan 
(small crosses and small upper limits).
The maximum contribution to the X-ray emission expected from CVs and normal stars
(solid line) and from LMXBs (dotted line) is plotted as well. 
The AWM~5 group members follow the distribution of early-type galaxies and are well above
the contribution expected from CVs and stars. All the group members with detected
X-ray emission, except D1, are also not consistent with emission coming exclusively from
LMXBs.
\label{ewan}}
\end{figure}

We compared the X-ray and infrared luminosities of the galaxies in AWM~5 with 
a bright nearby sample of early-type galaxies (Ellis \& O'Sullivan 2006). 
The X-ray luminosity (where there is no detection) was computed from the
background level detected in a $3^{\prime\prime}$ circle centered on the 
optical position
of each galaxy, after modeling out the general group emission at that radius
(from the $\beta$ model derived in \S~\ref{radanal}). 
Following Poisson's formula we estimated the number of counts
that would be necessary to get a 3-$\sigma$ detection to
obtain a flux and therefore a luminosity.
Following a procedure similar to that adopted by Revnivtsev et al. (2008) in the case of the 
unresolved X-ray emission in NGC~3379, in Figure~\ref{ewan} we plotted the $K_s$-band 
luminosities (expressed in solar units) together with the ratio of the x-ray luminosity
over the $K_s$-band luminosity, for both
the galaxies in AWM~5 and the sample of Ellis \& O'Sullivan (2006).
A reference value for the maximum expected contribution to the X-ray emission
coming from CVs and normal stars (solid line)
and from low mass X-ray binaries (LMXBs, dotted line) is shown in the plot as well.
The AWM~5 members follow the distribution of early-type galaxies from the bright sample. 
This is true also for the two galaxies D2 and D3, assuming they are group members.
It is clear from this plot that CVs and stars cannot account for the X-ray emission 
in any of the AWM~5 members.
While the X-ray emission from D1 could be due entirely to unresolved LMXBs,
the X-ray emission from D2, D3, D4 and D5 is clearly of nuclear origin. Based on their
upper limits we cannot exclude that also the other group members with undetected X-ray 
emission could harbor a nuclear source. 

\subsection{Black hole masses and Bondi accretion rates}\label{bondi}

In the cores of early-type galaxies, stellar mass loss is of the order of a few $10^{-5}$ up to 
a few $10^{-4}$ M$_\odot$ yr$^{-1}$ (e.g. Soria et al. 2006) and in most cases represents the most important
fueling source for their central black holes (usually larger than the fuel available from the hot gas).
For the AWM~5 group members, we estimated the size of the black hole and the amount of fuel available from 
the X-ray emitting gas following the $L_{K_s,bulge}$-$M_{BH}$ relation of Marconi \& Hunt (2003;
as done in \S~\ref{radio} for NGC~6269).
Assuming that all the $K_s$ luminosity in our galaxies comes from a bulge, we 
estimated $M_{BH}$ for all the AWM~5 members. The values of $M_{BH}$ are presented
in Table~\ref{mbh}, together with the corresponding ratio between the observed X-ray luminosity $L_X$
(which can be considered of nuclear origin in all galaxies except for D1)
and the expected Eddington luminosity $L_{Edd}$ for a black hole of mass $M_{BH}$. 
For our sample of AWM~5 group members, $L_X/L_{Edd}<5\times10^{-6}$ showing that all 
the black holes are accreting at a highly sub-Eddington rate.
Considering the electron density profile of the X-ray emitting gas 
(as determined in \S~\ref{physpar}) and the temperature profile of the gas, we estimated the 
accretion rate on the SMBH, from the Bondi (1952)
formula:
\begin{equation}
\dot{M}_{Bondi} = 1.66 \times 10^{-7} \;
M_7^2\;\;T_{2}^{-3/2} \;n_{0.1} \; \hbox{$\thinspace 
M_\odot \; yr^{-1}$} 
\end{equation}
where $M_7$ is the $M_{BH}$ in units of $10^7\; M_{\odot}$, $T_{2}$ is the
temperature in units of 2 keV and $n_{0.1}$ is the density in units
of 0.1 cm$^{-3}$.
Assuming an accretion efficiency of $\eta=0.1$ we derived the
corresponding Bondi luminosity.
However, it is worth noting that we do not have any information on the line of sight position
of the galaxies, therefore both the Bondi accretion rate and its associated 
luminosity can only be estimated
as upper limits, apart from the case of NGC~6269. Another caveat to take into consideration is 
that motion of the galaxies through the group gas reduces the accretion rate. 
Allowing for uncertainties in the accretion rate, it is quite 
clear that accretion from group hot gas is not important compared to stellar mass loss 
($>$ a few $10^{-5}$ M$_\odot$ yr$^{-1}$; e.g. Soria et al. 2006) in the AWM~5 
group members.


\section{Re-heating mechanisms and thermal conduction suppression in the cooling-core of AWM~5}
\label{reheat}

As shown in Fig.~\ref{tc_entr}, the gas cooling time within the central 30 kpc 
of AWM~5 is less than the Hubble time. However, this group harbors a
small dense $\sim8$~kpc CC but lacks a group-sized CC which 
is found in other relaxed
groups (e.g., NGC~1550, Sun et al. 2003; MKW 4, O'Sullivan et al. 2003). 
Also ``fossil'' groups, which are believed to be old and relaxed systems, are expected 
that a group-sized CC should develop in the center in a relatively short time
(i.e. a few Gyr).
Examples of ``fossil'' groups presenting a smaller CC
than expected from the cooling time derived in their central regions are indeed 
ESO306170 (Sun et al. 2004) and NGC 6482 (Khosroshahi et al. 2004).

The lack of a group-sized CC in AWM~5 can be reconciled
with the current age of the system, if an existing CC has been
re-heated. 
Three viable sources for reheating CCs in groups and clusters
are energy from mergers (which drive gas motions in the core), energy 
input from SMBH outbursts in the central galaxy and
thermal conduction. 

\subsection{Thermal conduction}\label{conduct}

While conduction may be effective at larger radii (e.g., 
Narayan \& Medvedev 2001; Zakamska 
\& Narayan 2003), it is not effective within all cluster (and group) cores. 
Since CC clusters have a positive temperature gradient, heat 
can be conducted inward toward the center, balancing the radiative cooling 
and arresting or retarding the formation of the CC
(e.g. Stewart et al. 1984; 
Bregman \& David 1988; 
Brighenti \& Mathews 2002), although this is usually an unstable process.

Gas with temperature $T$, density $n$, 
and a volume $V$ put in thermal contact with an
ambient gas having temperature $T+\Delta T$, over a contact surface having
an area $A$ will eventually reach thermal equilibrium with the surrounding gas
in a time $\tau$ which can be estimated as (from B\"ohringer \& Fabian 1989,
adapted to the 3-D case by e.g. Simionescu et al. 2008):
\begin{equation}
\tau \approx \frac{\Delta H}{qA} 
\approx \frac{\frac{5}{2}nk\Delta T }{f k_0 T^{5/2}\nabla T}\frac{V}{A}
\end{equation}
where $\Delta H$ is the enthalpy that needs to be transferred to 
the lower temperature gas to reach thermal equilibrium, $q$ is the
conductive heat flux, $f$ is a reduction coefficient that takes into account
suppression of heat conduction by magnetic fields, and 
$k_0\sim6\times10^{-7}$~erg cm$^{-1}$ s$^{-1}$ K$^{-7/2}$
is the Spitzer conduction coefficient (Spitzer 1962).
For AWM~5, we have a gas temperature $kT=1.2$~keV in a sphere
with an $\sim8$~kpc radius, in contact with ambient gas with a temperature of 3.7~keV. 
For $f=1$, a density $n=0.05$~cm$^{-3}$ and a temperature gradient which is the
difference in temperature between the core and the bin right outside the core divided
by the distance between the center of the two bins, we obtain a very short 
timescale of $\sim17\pm3$~Myr to reach thermal equilibrium between the two gases. 
Moreover, the conductive heat flux is $\sim7\times10^{43}$
erg s$^{-1}$, about $\sim100$ times larger than the luminosity emitted in X-rays
from the core. Both of these facts require a large suppression of the thermal
conductivity in the core of AWM~5, by a factor of at least $\sim100$. 
Such a large suppression factor is predicted by some theoretical calculations (e.g. Rechester \& Rosenbluth
1978; Chandran \& Cowley 1998) and supported by observations in other clusters and groups
of galaxies (e.g. Coma; Vikhlinin et al. 2001b). This is also consistent with the results of Simionescu
et al. (2008) for M87. 


\subsection{Gas sloshing}\label{slosh}

``Sloshing'' of the
gas in the group core, generated by
past merger activity, could provide
enough energy to heat the gas in this region
(see e.g., Churazov et al. 2003 for the case of Perseus). 
Markevitch et al. (2003b) found that approximately 2/3 of relaxed, 
centrally peaked clusters with CC showed sharp surface 
brightness discontinuities or ``edges'', characteristic of gas motions. 
After accounting for clusters with ``edges'' not favorably oriented for detection, 
all CC clusters likely harbor such features. 
Ascasibar \& Markevitch (2006) argue that in centrally 
peaked systems, ``edges'' (or cold fronts) are driven by mergers and that
merger driven ``sloshing'' is long lived since the
dark matter does not relax quickly. 
However, signatures of such merger activity (e.g. surface brightness ``edges'') are not
clearly visible in the X-ray morphology of AWM~5. The only discontinuities visible
in the surface brightness profile are of very low statistical significance and even in the
azimuthal profiles the discrepancies from a $\beta$ model are within the errors.
This explanation cannot be accurately tested with the present data and would need a deeper
observation to confirm or rule out the ``sloshing'' as a possible re-heating source
of the CC of AWM~5.

\subsection{Central SMBH outbursts}\label{outburst}

Chandra observations have shown buoyant bubbles and weak 
shocks around central cluster galaxies powered by AGN 
outbursts (e.g. Perseus, Fabian et al. 2003; M87, Forman et al. 2005). 
Churazov et al. (2001, 2002) argued that cyclical outbursts from 
self-regulated accretion onto the central black hole could be the source of 
energy to (re)heat the CCs.
Thus, a central AGN outburst might provide the required energy to heat the outer regions
between 8~kpc and 30~kpc. 
Although the presence of a central AGN is not directly observed in the X-ray data 
($L_X<5.4\times 10^{40}$ erg s$^{-1}$), it is very likely that NGC~6269
harbors a now dormant $10^9$ solar mass black hole (from its large $K_s$ band 
luminosity). 
As described in \S~\ref{radio} a weak radio source is present at the center of 
NGC~6269 with a total flux (core plus lobes) of $\sim45$ mJy at 1400 MHz. 
Assuming the spectral shape in the radio derived by Giacintucci et al. (2007), 
we can compute a total radio luminosity (10
MHz - 10 GHz) of $\sim$9 $\times 10^{39}$ ergs
s$^{-1}$. Taking the relation 
of B\^{\i}rzan et al. (2004), radio sources with this total 
luminosity can have  
mechanical luminosities up to 2$\times$10$^{43}$ ergs s$^{-1}$. 
If the central AGN has been active for the past 10$^{8}$ yr (consistent
with the age of the radio source derived in \S~\ref{radio} from the extension
of the lobes and with the radiative age determined by Giacintucci et al. 2007), 
we need a heating rate of $\sim$3$\times 10^{43}$ ergs s$^{-1}$, consistent
with the 2$\times$10$^{43}$ ergs s$^{-1}$ obtained from the radio properties of 
the group.
Therefore, it is
possible that a central AGN outburst could be partly responsible for re-heating the gas 
between 8 and 30~kpc. A signature of this process may also be seen in the
entropy profile (Fig.~\ref{scalentr}). The lower entropy gas between 8 and 30~kpc
could have been removed by the AGN outburst (e.g. Br\"uggen \& Kaiser 2002).
Moreover, the lifetime of
the radio synchrotron emission is only 24 (B/10 $\mu$G)$^{-3/2}$
($\nu$/1.4 GHz)$^{-1/2}$ Myr after the last injection of relativistic electrons
from the nucleus. 
In this scenario the energy transported by the radio jets was deposited beyond
8~kpc (consistent with the observed extension of the radio lobes) and heated the
CC only beyond that radius.
Weak shocks could maintain the temperature of the gas at smaller radii (see e.g. Forman et al. 2007,
McNamara \& Nulsen 2007, for recent examples).

\section{The ram-pressure stripped tail in NGC~6265}\label{6265}

\begin{figure}[h]
\begin{center}
\plotone{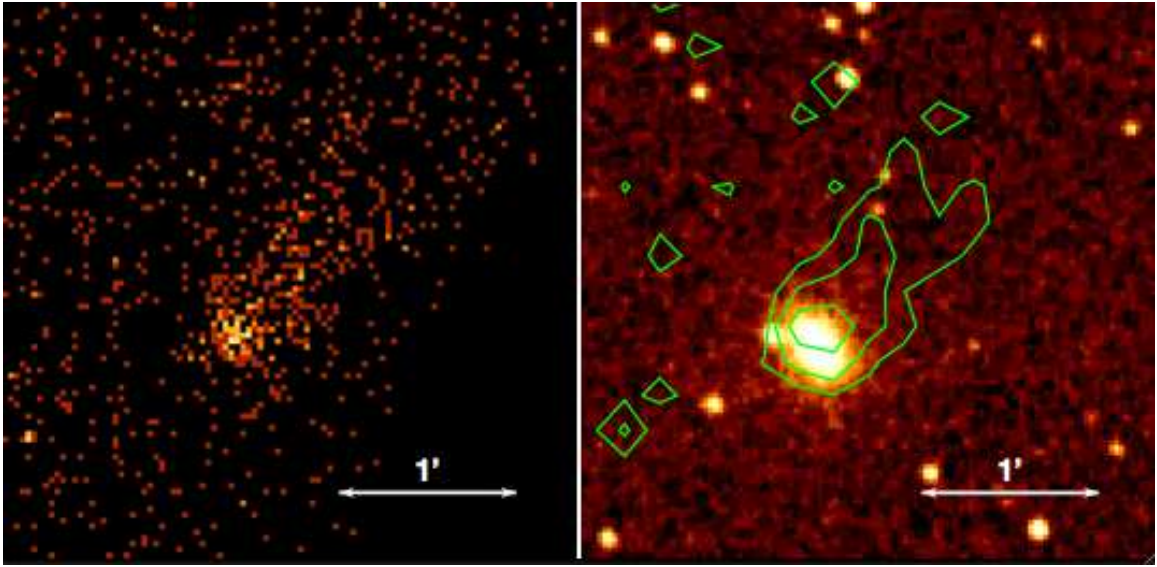}
\caption{{\it Left:} X-ray emission (0.5-2 keV; 
4$^{\prime\prime}$ bins) of ram pressure stripped gas from NGC~6265.
{\it Right:} X-ray contours superimposed on the optical DSS image. The
X-ray tail is 40~kpc long (projected on the sky).}
\label{ngc6265}
\end{center}
\end{figure}

Fig.~\ref{ngc6265} shows signs of possible ram pressure stripping of the atmosphere 
from the group member NGC~6265, an S0 galaxy. The trailing stream of hot gas is at least 
40 kpc long (in the plane of the sky). 

\begin{figure}[h]
\begin{center}
\plotone{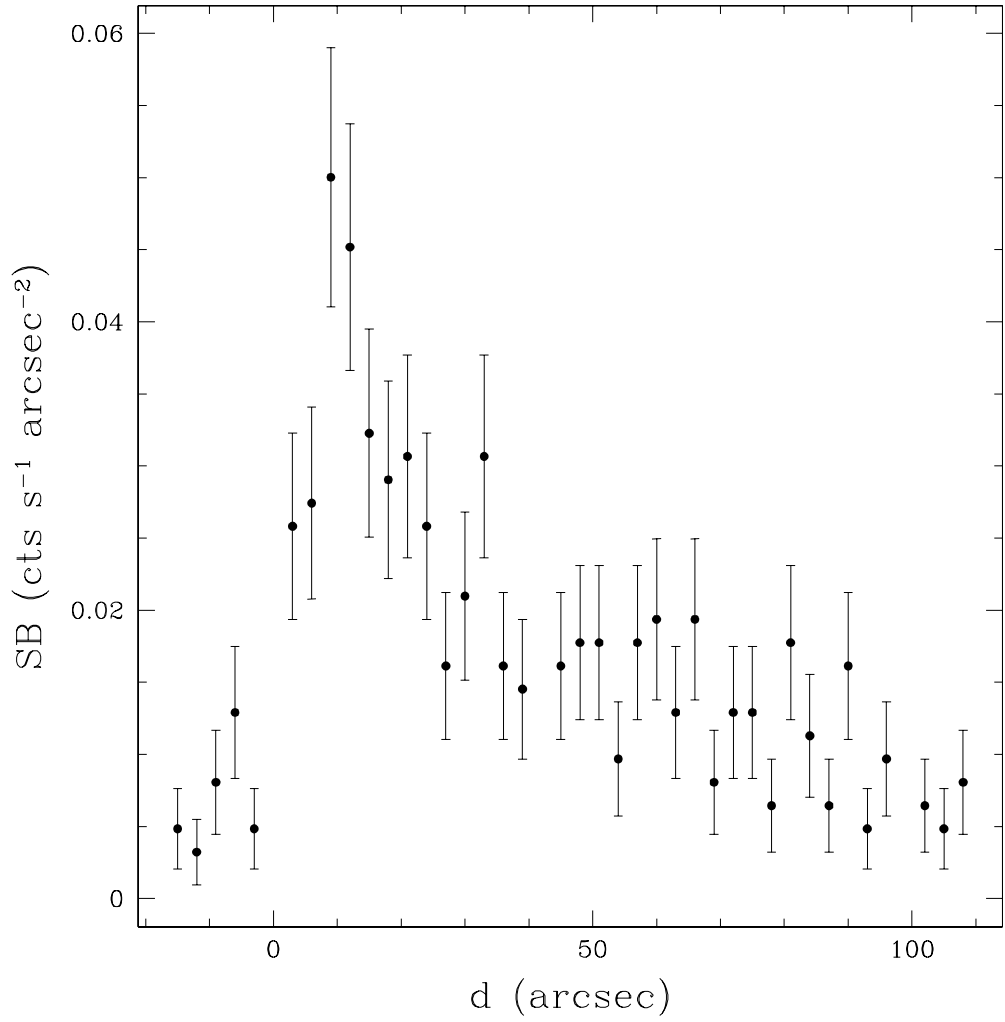}
\caption{Surface brightness profile along the X-ray gas stream around NGC~6265. The zero on the X-axis
corresponds to the leading edge of the X-ray emission, where a sudden drop in surface brightness
of a factor of 3-4 is observed.}
\label{sb6265}
\end{center}
\end{figure}

The surface brightness profile (Fig.~\ref{sb6265}) shows a sudden drop (of a factor of $3-4$ at least) 
in coincidence with the leading edge of the x-ray emission, while going toward the tail the decrease
in surface brightness is definitely more gradual. 
We extracted spectra in two different regions in the galaxy, one coincident
with the core and a rectangular region coincident with the tail.
We fit a thermal $mekal$ model to the core of the galaxy, obtaining a temperature
$kT=0.78_{-0.08}^{+0.12}$ keV (with Z fixed at 0.3 solar). 
For the tail, we measure $kT=0.63_{-0.07}^{+0.09}$ keV ($Z\equiv0.3Z_\odot$). 
A second thermal component (with $kT\equiv1.8$~keV, as measured from the temperature
profile at the radial distance of NGC~6265) was necessary to take into account the contribution from
the ICM of AWM~5. The gas temperature observed in the tail is consistent with the temperature present
in the core, and more than $\sim2$ times lower than the AWM~5 gas at the radial distance of the galaxy, 
supporting a scenario where this is predominantly ram pressure stripped gas coming 
from NGC~6265 instead of ICM gas concentrated in the wake.
We followed Vikhlinin et al. (2001a) to determine the speed of NGC~6265 with respect to the ambient
ICM. The ratio of the pressure $p_0$ 
measured at the stagnation point (at the edge of the discontinuity) to the pressure $p_1$
measured in the free-stream, i.e. in the undisturbed AWM~5 ICM, is directly related to the Mach
number $M$ of the NGC~6265 motion relative to the ICM (Landau \& Lifschitz 1959; Vikhlinin et al. 2001a).
Although the gas pressure at the stagnation point cannot be measured directly, it must equal the 
pressure within the leading edge. We assume indeed that the pressure at the stagnation point is
equal to the pressure measured in a region including the core of the galaxy, 
i.e. $p=5.2\pm1.0\times10^{-11}$~dyn 
cm$^{-2}$. 
A more precise estimate would be to measure the pressure in a thin arc located along the leading
edge. However, the counts are too few to perform such a measure. 
The ratio to the pressure measured in the ICM is $p_0/p_1=17.3_{-5.3}^{+5.6}$, which
corresponds to a Mach number $M=3.4_{-0.6}^{+0.5}$. $M\ga3$ requires the presence
of a shock front with a density jump of a factor of 3. The surface brightness profile observed in NGC~6265
shows a drop corresponding to the leading edge, consistent with such a scenario (Fig.~\ref{sb6265}).
We derive that the total velocity
at which NGC~6265 is moving relatively to the ICM is $v\sim2300$~km s$^{-1}$. This very high
3D velocity is a further confirmation that NGC~6265 is not a resident of the AWM~5 core (\S~\ref{intro})
Based on the redshift of AWM~5 and NGC~6265 
the velocity along the line of sight is $\sim700$~km s$^{-1}$, therefore the component
of the velocity parallel to the plane of the sky is $\sim2200$~km s$^{-1}$ and the inclination
angle of the motion of the galaxy with respect to the plane of the sky is $\xi\sim18^\circ$. 
The physical length of the tail should be
then $\sim42$~kpc and the ram pressure exerted by the ICM on the NGC~6265 gas because of the supersonic
motion through it is $p_{ram}\sim4\times10^{-11}$~dyn cm$^{-2}$. 
Unfortunately the number of photons detected in this observation
(about 300 in the core and 
in the tail combined) 
is not sufficient for a detailed analysis.
However, we derive the mean gas density, assuming that the tail is a cylinder with an $18^\circ$ 
inclination to the plane of the sky. The gas density derived from the spectral analysis is
$n_e=4.1\pm0.3\times10^{-3}$~cm$^{-3}$, at least one order of magnitude denser than the ICM gas
($n_{e,ICM}<3\times10^{-4}$~cm$^{-3}$). The mass of the tail, derived assuming cylindrical geometry, is
$2.1\pm0.2\times10^9$~M$_\odot$.

\section{Discussion}

As discussed in \S~\ref{reheat}, AWM~5 is harboring a
dense $\sim8$~kpc sized CC, smaller 
than expected from the cooling time derived in its central regions.
The required energy to heat the 1.2~keV gas (observed in the core at $r<8$~kpc) 
between 8 and 30 kpc to the observed
value ($kT\sim3.7$~keV) is estimated to be $\sim10^{59}$ ergs.

Such a large amount of energy cannot be clearly produced by star formation activity
in the central galaxy NGC~6269 which is a cD with practically no ISM detected (e.g. 
Bettoni et al. 2003). 
SNIa explosions could be in principle another heating source. 
If we consider
the average supernova rate observed in elliptical galaxies (Turatto et al. 1994)
rescaled to the B-band luminosity of NGC~6269 we obtain a rate of 0.89 SN every 100 yr.
Considering that each supernova could produce an energy up to $10^{51}$~ergs we obtain a 
corresponding heating rate of $\la 2.8\times10^{41}$~erg s$^{-1}$. To obtain a total energy
of  $\sim10^{59}$ ergs we would have need a constant supernova rate for more than 10~Gyr.
Therefore we can affirm that the contribution of SN to the heating of the gas around
the AWM~5 core is negligible.
As explained in \S~\ref{slosh}, signatures of ``sloshing'' of the
gas in the group core, generated by past merger activity
(e.g. surface brightness ``edges'') could not be found in the X-ray observations of AWM~5. 
Only deeper data could provide a meaningful test of the importance of sloshing as a
re-heating source for the CC of AWM~5.

The only two sources which could have played an important role in the CC re-heating are
the thermal conduction and the central SMBH outburst.
The observed temperatures in the center of AWM~5 yields a short timescale of $\sim17$~Myr 
to reach thermal equilibrium between the lower temperature gas in the core and the higher
temperature gas between 8 and 30~kpc.  
Moreover, the heat flux expected corresponds to $L\sim7\times10^{43}$
erg s$^{-1}$, about $\sim100$ times larger than the luminosity emitted in X-rays
from the core ($\sim5\times10^{41}$
erg s$^{-1}$). It is clear that these two gases cannot stay in the current equilibrium 
situation if a large suppression of the thermal conductivity in the core of AWM~5
(by a factor of at least $\sim100$) is not present. 
Large suppression factor are indeed predicted by theory (e.g. Chandran \& Cowley 1998) 
and observed in galaxy clusters and groups (e.g. Coma; Vikhlinin et al. 2001b).
The most likely scenario is that thermal conduction has been suppressed by a large 
factor (most likely because of the presence of magnetic fields) and did not contribute
significantly to CC re-heating.
A central AGN is not directly observed in the AWM~5 X-ray data 
($L_X<5.4\times 10^{40}$ erg s$^{-1}$). However the $K_s$ band luminosity suggests that
NGC~6269 is harboring a now dormant $10^9$ solar mass black hole.
As a further confirmation to this scenario, a weak radio source is detected
in the center of NGC~6269 (see \S~\ref{radio}). 
As estimated in \S~\ref{outburst} a radio source with a total radio luminosity (10
MHz - 10 GHz) of $\sim$9 $\times 10^{39}$ ergs
s$^{-1}$ could have  
mechanical luminosities up to 2$\times$10$^{43}$ ergs s$^{-1}$ (B\^{\i}rzan et al. 2004). 
To produce the $\sim10^{59}$ ergs necessary to reheat the gas outside the core it is
necessary that the AGN was active for the past $\ga10^{8}$ yr (consistent
with the radiative age determined by Giacintucci et al. 2007).
Therefore, it is very likely that a central AGN outburst could be the main responsible 
for re-heating the gas between 8 and 30~kpc from the center, showing its signatures on the
entropy profile (Fig.~\ref{scalentr}). Indeed, the lower entropy gas between 8 and 30~kpc
may have been removed by the AGN outburst (e.g. Br\"uggen \& Kaiser 2002).

\section{Conclusions}

In this paper we presented an analysis of a 40~ksec Chandra observation of the 
galaxy group AWM~5 (Albert et al. 1977). Although it does not meet the formal definition of
a fossil group given in Jones et al. (2003), this group is a system with a dominant 
central galaxy, making it well suitable for a comparison with
both ``fossil'' and normal groups.
The main results from our analysis are:

\begin{itemize}
\item the abundance distribution shows a ``hole'' at the center, coincident with a CC. 
The maximum value
of the abundance is at least a few times solar (at $\sim15-20$ kpc from
the center). The presence of an abundance ``hole'' is not 
uncommon in groups and clusters of galaxies
(e.g. Perseus cluster; Schmidt et al. 2002; Churazov et al. 2001; 2002)
and can be suggestive of metals being driven from the core;
\item the derived deprojected electron density profile can be fit by
two $\beta$-models. The inner one (at $r<13.4$~kpc)
has $\beta=0.72_{-0.11}^{+0.16}$ and $r_c=5.7_{-1.5}^{+1.8}$~kpc). At  $r>13.4$~kpc
we derived $\beta=0.34\pm0.01$ and $r_c=31.3_{-5.5}^{+5.8}$ kpc.
At $r>15$~kpc a power-law is also a good fit to the data with 
n$_{\rm e} \propto$ r$^{-0.91 \pm 0.02}$ between 15 kpc and 300 kpc,
a similar slope to that observed in cool groups
(e.g. NGC1550, Sun et al. 2003) but much flatter than those measured in ``fossil'' groups
(e.g. NGC6482, Khosroshahi et al. 2004; ESO3060170, Sun et al. 2004) and in clusters;
\item we compared the X-ray and optical properties of the group members inside the
Chandra field of view. 
Apart from NGC~6269 and NGC~6265 
only three of the 12 group galaxies in the Chandra field of view were detected. Two
other X-ray sources within 1$^\prime$ ($\sim40$~kpc) of NGC~6269 have red, galaxy-like
optical counterparts and are likely group members. 
We estimated $M_{BH}$ for the group members from their $K_s$ luminosity following
the relation of Marconi \& Hunt (2003). The ratio $L_X/L_{Edd}$ is less than 
$5\times10^{-6}$ showing that these black holes are accreting
at highly sub-Eddington rates.
\item we determined that the gas cooling time within the central 30 kpc of AWM~5 is smaller 
than the Hubble time, but only a small ($r\sim8$~kpc) CC is observed. 
This discrepancy suggests that the gas beyond $\sim8$~kpc has been
re-heated;
\item we find that heat conduction could be a significant source of re-heating in the AWM~5 core.
Moreover, balance between heating expected from conduction and cooling due to 
X-ray emission requires large suppression (of a factor of at least
$\sim100$) in the conductivity;
\item we find evidence for past AGN activity 
(whose footprints are still observable as a radio 
source in the center of the cD galaxy NGC~6265). This is the most likely contributing source 
of re-heating for the central regions of AWM~5. 
\item we studied the properties of the ram pressure stripped tail in the group member NGC~6265.
The galaxy is moving at $M\approx3.4_{-0.6}^{+0.5}$ in the group ICM. Moreover
the physical length of the tail was measured to be $\sim42$~kpc.
\end{itemize}

\acknowledgments
We are grateful to M. Sun for providing us with the entropy profiles of the groups in
his sample. We thank S. Giacintucci and M. Markevitch for useful discussions.
We thank the anonymous referee for useful suggestions that helped to improve the
presentation of the results in the paper.

\clearpage



\clearpage

\clearpage

\clearpage

\clearpage


\clearpage

\begin{deluxetable}{cccc}
\tabletypesize{\normalsize}
\tablecaption{Best fit $\beta$ models for the surface brightness profile wedges of AWM~5}
\tablewidth{0pt}
\tablehead{
\colhead{Wedge Orientation} &
\colhead{$\chi^2$/$dof$} &
\colhead{$\beta$} & 
\colhead{$r_c$ (kpc)}
}
\startdata
East & 37.7/26 & $0.352\pm0.005$ & $19.7_{-5.3}^{+5.9}$ \\
North & 30.8/26 & $0.383\pm0.006$ & $24.7_{-4.6}^{+5.1}$ \\
South & 34.2/26 & $0.347\pm0.005$ & $7.0_{-3.1}^{+3.8}$ \\
West & 29.9/26 & $0.359\pm0.005$ & $9.4_{-3.6}^{+3.2}$ \\
\enddata
\label{wedges}
\end{deluxetable}

\clearpage

\begin{landscape}
\begin{deluxetable*}{ccccccccccccc}
\tabletypesize{\scriptsize}
\tablecaption{Galaxies with velocities consistent with the AWM~5 group 
within the Chandra ACIS-I field of view and relative
optical-infrared properties. The optical colors are taken from the third data release of SDSS
(Abazajian et al. 2005), while the infrared are from 2MASS. 
The radial velocities are from Koranyi \& Geller (2002).
}
\tablewidth{0pt}
\tablehead{
\colhead{\#} &
\colhead{\begin{tabular}{c}
RA\\ 
(J2000)
\end{tabular}} &
\colhead{\begin{tabular}{c}
Dec\\ 
(J2000)
\end{tabular}} & 
\colhead{$u$} & 
\colhead{$g$} & 
\colhead{$r$} & 
\colhead{$i$} & 
\colhead{$z$} & 
\colhead{B} & 
\colhead{\begin{tabular}{c}
log $L_B$\\ 
($L_\odot$)
\end{tabular}} & 
\colhead{$K_s$} & 
\colhead{\begin{tabular}{c}log $L_{K_s}$\\
($L_\odot$)
\end{tabular}} & 
\colhead{\begin{tabular}{c}
$v_{r}$\\
(km s$^{-1}$)
\end{tabular}} }
\startdata
N6269 & 16:57:58.1 & +27:51:16 & 15.5 & 13.5 & 12.6 & 12.1 & 11.7 & 14.1 & 10.87 &  9.40 & 11.91 &   10460 \\
N6265 & 16:57:29.1 & +27:50:39 & 16.7 & 14.7 & 13.8 & 13.4 & 13.0 & 15.3 & 10.39 & 10.75 & 11.37 &    9725 \\
D1 & 16:58:44.1 & +27:51:32 & 17.2 & 15.2 & 14.3 & 13.9 & 13.6 & 15.8 & 10.19 & 11.35 & 11.13 &    9720 \\
D2 & 16:57:55.9 & +27:51:52 & 20.0 & 17.5 & 16.1 & 15.0 & 14.5 & 18.3 &  9.19 & 12.40 & 10.71 &  (10670)\\
D3 & 16:57:55.4 & +27:52:01 & 20.1 & 17.6 & 16.2 & 15.1 & 14.5 & 18.4 &  9.15 & 12.45 & 10.69 &  (10670)\\
D4 & 16:57:54.6 & +27:52:17 & 18.8 & 16.8 & 16.0 & 15.5 & 15.2 & 17.3 &  9.59 & 13.00 & 10.47 &   10670 \\
D5 & 16:57:49.2 & +27:55:06 & 18.6 & 16.9 & 16.1 & 15.7 & 15.4 & 17.4 &  9.55 & 13.75 & 10.17 &   10371 \\
U1 & 16:58:40.8 & +27:53:10 & 18.3 & 16.5 & 15.6 & 15.1 & 14.8 & 17.0 &  9.71 & 12.76 & 10.56 &   10722 \\
U2 & 16:58:19.4 & +27:57:35 & 18.3 & 16.3 & 15.4 & 14.9 & 14.6 & 16.9 &  9.75 & 12.24 & 10.77 &    9976 \\
U3 & 16:58:18.3 & +27:48:30 & 18.5 & 16.7 & 15.8 & 15.4 & 15.0 & 17.3 &  9.59 & 13.05 & 10.45 &   10194 \\
U4 & 16:57:55.1 & +27:41:58 & 18.0 & 16.1 & 15.2 & 14.7 & 14.4 & 16.7 &  9.83 & 12.44 & 10.69 &   10977 \\
U5 & 16:57:53.8 & +27:56:18 & 18.2 & 16.4 & 15.5 & 15.1 & 14.8 & 16.9 &  9.75 & 12.73 & 10.58 &   11510 \\
U6 & 16:57:48.4 & +27:55:20 & 19.0 & 17.3 & 16.4 & 16.0 & 15.7 & 17.9 &  9.35 & 13.59 & 10.23 &   11159 \\
U7 & 16:57:34.7 & +27:48:28 & 18.6 & 16.8 & 15.9 & 15.5 & 15.2 & 17.4 &  9.55 & 12.74 & 10.57 &   11088 \\
U8 & 16:57:33.5 & +27:55:50 & 18.9 & 16.9 & 16.0 & 15.5 & 15.2 & 17.5 &  9.51 & 13.26 & 10.36 &   11237 \\
U9 & 16:57:21.3 & +27:53:26 & 19.1 & 17.3 & 16.4 & 16.0 & 15.7 & 17.9 &  9.35 & 13.71 & 10.18 &   10822 \\
\enddata
\label{optical}
\end{deluxetable*}

\clearpage
\end{landscape}

\begin{landscape}
\begin{deluxetable*}{crrrrr}
\tabletypesize{\footnotesize}
\tablecaption{Black hole masses, X-ray luminosities and Bondi accretion rates for the AWM~5 group members
in the ACIS-I field of view. The black hole masses were derived from $L_{K_s}$ following Marconi \& Hunt (2003). 
The Bondi accretion rate (Bondi 1952) was determined assuming the temperature and density of the gas at 
the position of each galaxy as derived in \S~\ref{physpar} and therefore can be considered upper limits
apart from the case of the central galaxy NGC~6269. 
The Bondi luminosities were computed
assuming an accretion efficiency $\eta=0.1$.
}
\tablewidth{0pt}
\tablehead{
\colhead{\#} &
\colhead{\begin{tabular}{c}$M_{BH}$\\ 
($M_\odot$)
\end{tabular}} & 
\colhead{\begin{tabular}{c}log $L_{0.3-10\:keV}$\\ 
(erg s$^{-1}$)
\end{tabular}} & 
\colhead{$L_{0.3-10\:keV}/L_{Edd}$} & 
\colhead{\begin{tabular}{c}
$\dot{M}_{Bondi}$\\
(M$_\odot$ yr$^{-1}$)
\end{tabular}} &
\colhead{\begin{tabular}{c}
log $L_{Bondi}$\\
(erg s$^{-1}$)
\end{tabular}} }
\startdata
N6269 &  $1.6\times10^9$ & $ <40.73$ & $ 2.6\times10^{-7}$ & $6.0\times10^{-3}$ & $43.53$ \\
N6265 &  $6.0\times10^8$ & $ <40.35$ & $ 2.9\times10^{-7}$ & $<5.7\times10^{-7}$ & $<39.51$ \\
D1 &  $2.8\times10^8$ & $ 39.91$ & $ 2.3\times10^{-7}$ & $<1.3\times10^{-4}$ & $<41.87$ \\
D2 &  $         10^8$ & $ 40.12$ & $          10^{-6}$ & $<1.7\times10^{-5}$ & $<40.97$ \\
D3 &  $         10^8$ & $ 40.01$ & $          10^{-6}$ & $<1.7\times10^{-5}$ & $<40.97$ \\
D4 &  $  6\times10^7$ & $ 40.04$ & $ 1.4\times10^{-6}$ & $<5.9\times10^{-6}$ & $<40.53$ \\
D5 &  $2.5\times10^7$ & $ 40.22$ & $   5\times10^{-6}$ & $<1.0\times10^{-6}$ & $<39.77$ \\
U1 &  $6.3\times10^7$ & $<40.11$ & $<1.6\times10^{-6}$ & $<6.6\times10^{-6}$ & $<40.57$ \\
U2 &  $1.1\times10^8$ & $<39.94$ & $<6.2\times10^{-7}$ & $<2.0\times10^{-5}$ & $<41.06$ \\
U3 &  $  6\times10^7$ & $<40.05$ & $<1.4\times10^{-6}$ & $<6.0\times10^{-6}$ & $<40.53$ \\
U4 &  $         10^8$ & $<40.03$ & $<8.2\times10^{-7}$ & $<1.7\times10^{-5}$ & $<40.97$ \\
U5 &  $6.3\times10^7$ & $<40.14$ & $<1.7\times10^{-6}$ & $<6.6\times10^{-6}$ & $<40.57$ \\
U6 &  $3.2\times10^7$ & $<39.95$ & $<2.1\times10^{-6}$ & $<1.7\times10^{-6}$ & $<39.98$ \\
U7 &  $6.3\times10^7$ & $<39.83$ & $<8.2\times10^{-7}$ & $<6.6\times10^{-6}$ & $<40.57$ \\
U8 &  $  4\times10^7$ & $<39.97$ & $<1.8\times10^{-6}$ & $<2.7\times10^{-6}$ & $<40.18$ \\
U9 &  $2.5\times10^7$ & $<39.93$ & $<2.6\times10^{-6}$ & $<1.0\times10^{-6}$ & $<39.77$ \\
\enddata
\label{mbh}
\end{deluxetable*}
\clearpage
\end{landscape}


\end{document}